# Tailoring the ultrafast dynamics and electromagnetic surface modes on silicon upon irradiation with mid-infrared femtosecond laser pulses using SiO2 coatings

G. D. Tsibidis [1♣] and E. Stratakis [1,2*]

[1] *Institute of Electronic Structure and Laser (IESL), Foundation for Research and Technology (FORTH), N. Plastira 100, Vassilika Vouton, 70013, Heraklion, Crete, Greece*

[2] *Department of Physics, University of Crete, 71003 Heraklion, Greece*

A key issue in the use of high-power mid-infrared (Mid-IR) laser sources for a plethora of applications is the investigation of the exciting laser driven physical phenomena taking place in materials coated with dielectric films. Here, we present a theoretical investigation of the ultrafast processes and thermal response upon excitation of two-layered complexes consisting of fused silica thin films placed on silicon substrates with ultrashort pulsed lasers in the Mid-IR spectral regime. Through the development of a theoretical model, we demonstrate that the control of the underlying ultrafast phenomena and the damage threshold (DT) of the substrate are achieved via an appropriate modulation of the thickness of the $SiO_2$ film. It is shown that a decrease of DT by up to ~30% compared to the absence of coating is feasible emphasising the impact of coatings of lower refractive index than the substrate. The conditions for surface plasmon (SP) excitation on the interface between $SiO_2$ and Si and the influence of the film thickness on the SP features are also discussed as such electromagnetic modes can initiate structuring on the interface. Our results manifested a striking impact of the presence and the coating thickness on the SP characteristics. It is shown that the SP excitation can be tailored by changing the superstrate thickness which also determines the competition of the air and the dielectric in the SP characteristics. These remarkable predictions can be employed for the development of new optical coatings and components for nonlinear optics and photonics for a large range of Mid-IR laser-based applications.

## I. INTRODUCTION

Irradiation of semiconducting materials (i.e. Silicon, Germanium, etc.) with high intensity femtosecond (fs) laser beams in the Mid-IR spectral region raises challenging opportunities in photonics and an abundance of applications [1-5] due to the exciting laser driven phenomena compared with pulses at lower wavelengths [6-8]. The pronounced transparency of the semiconductors at Mid-IR compared to their behaviour at the visible and near-infrared regimes (i.e. leading to a noticeable absorption dynamics [6-8]), the excitation of electromagnetic modes such as surface plasmon (i.e. crucial for laser machining purposes [9] and for functioning of the material as a novel plasmonic tool [4, 5, 10]) at lower excitation levels, and the expected influence of the electron excitation/plasma formation through the scaling of the ponderomotive energy $U_p$ with the laser wavelength (i.e. $U_p \sim \lambda_L^2$ where $\lambda_L$ stands for the laser wavelength) constitute some physical phenomena which are characteristic of the Mid-IR spectral region [3].

It is evident that a precise knowledge of the fundamentals of laser interaction with the target material and the elucidation of the aforementioned phenomena in various laser conditions are very crucial for the efficient employment of the Mid-IR based technology in various applications. In particular, a key technological and fundamental challenge in the research of how materials respond upon exposure to Mid-IR fs pulses is relevant to the determination of the damage threshold (DT) of the target which is defined as the smallest fluence that induces minimal damage on the surface of the irradiated solid. In previous reports, DT measurements [11, 12] and predictions [9, 11] were presented in various laser conditions and laser wavelengths for Silicon; it has been deduced that understanding of the fundamentals of the strong field interaction with Silicon (Si) [9, 13] or Germanium (Ge) [14] in the Mid-IR regime can lead to efficient laser-based patterning.

Nevertheless, despite the interesting physical phenomena produced on bulk semiconductors irradiated with strong laser fields, there still exist fundamental open questions in regard to the effects on two-layered semiconducting/dielectric materials. More specifically, it is evident that a lack of knowledge of how coated materials respond to Mid-IR pulses and how DT scales with $\lambda_L$ prevents the optimization of the use of pulses in this regime for technological applications. Given the transparency of low- and high- bandgap materials in the Mid-IR regime, Si, ZnS, Ge as well as fluorides and fused Silica ($SiO_2$) have been used as high-performance coatings for Mid-IR related applications [15, 16]. In a recent report [17], it was shown that the employment of Si films as a coating on top of $SiO_2$ *increases* the damage threshold of the bulk substrate and this behaviour is dependent on the thickness of Si. Furthermore, the optical parameters of the two-layered material can also be modulated via an appropriate selection of the coating thickness.

On the other hand, a key issue in the manufacturing of optical components providing the opto-thermal response of Si at long wavelengths, is whether a two-layered complex consisting of a transparent coating of *lower* refractive index than that of the semiconductor on top of bulk Si influences the behaviour of the substrate. Fused silica represents a well-characterised material and a fundamental question is pertinent to whether $SiO_2$ coatings of various thicknesses could result in a different opto-thermal response of the Si substrate upon irradiation with Mid-IR fs pulses. Thus, the evaluation of how a $SiO_2$/Si complex behaves can potentially lead to an increase of the throughput of the system, allow a control of the DT of the substrate, enable the reduction of undesired effects generated by reflections and cause subsurface laser-induced modifications. It is evident that such an assessment requires unraveling the ultrafast phenomena that take place during irradiation of $SiO_2$/Si with strong Mid-IR fs pulses. The elucidation of the aforementioned issues is of paramount importance not only to understand further the underlying physical processes of laser-matter interactions and ultrafast electron dynamics but also to associate the induced thermal effects with damage on the substrate. To this end, a systematic exploration of the correlation of the laser parameters (i.e. fluence, pulse duration, wavelength) and material features (i.e. thickness of film) with the resulting effects on the surface of the substrate are aimed to potentially create novel capabilities for material processing.

In regard to the capacity for material processing, previous reports showed that laser-induced structural modification with laser pulses can be confined inside the volume of a multi-layered stack comprising of dielectric materials of different refractive indices [18]. The fabrication of subwavelength laser-induced modifications inside the volume of the stack without the employment of tightly focused femtosecond laser beams offer a novel methodology for micro/nano-fabrication which can benefit a variety of applications, including metalenses, anticounterfeiting marking, waveguides, circuits and high-density data storage in durable materials [18-20]. Given the transparent character of both $SiO_2$ and Si (at Mid-IR), it is interesting to explore conditions to achieve similar high spatial confinement of laser-based modification inside a two-layered complex. By recalling that Surface Plasmon (SP) excitation constitutes one of the predominant candidate mechanisms that lead to the fabrication of Laser-Induced Periodic Surface Structures [21, 22], it is important both from a fundamental and application point of view to discuss whether SP excitation is attainable on the interface of a two-layered material complex irradiated with Mid-IR pulses.

Motivated by the above challenges, in the current work, a detailed theoretical investigation is conducted aiming to correlate the ultrafast dynamics, thermal effects, damage threshold and SP excitation on the Silicon substrate in various laser conditions and $SiO_2$ coating thicknesses. The paper is organised as follows: in Section II, a detailed theoretical framework is presented to describe the physical processes that occur upon irradiation of $SiO_2$/Si with Mid-IR fs laser and the simulation procedure is described in Section III. A systematic analysis of the results and discussion are illustrated in Section IV and V, respectively, while concluding remarks follow in Section VI.

## II. THEORETICAL MODEL

A key role in the elucidation of the effects that lead to material damage is the energy absorption from the two-layered material; in particular, the composition of the complex and thickness of the constituent materials are aimed to tailor the optical parameters and, therefore, determine the amount of absorbed energy. Thus, a crucial investigation requires the evaluation of the optical properties of $SiO_2$/Si structures for various thicknesses $d$ of the $SiO_2$ coating. In Fig.1, the two-layered complex, $SiO_2$/Si, irradiated with mid-IR pulses is sketched.

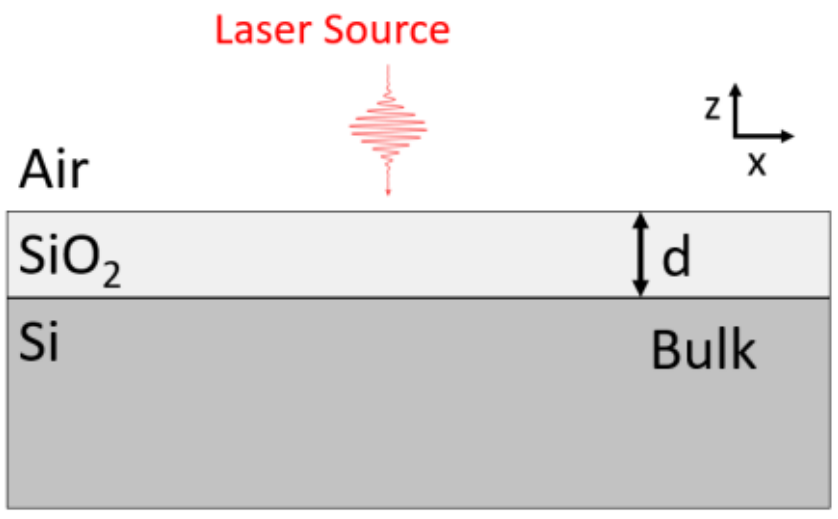

Figure 1: A two-layered complex (a $SiO_2$ film of thickness $d$ on top of bulk Si irradiated with a mid-IR laser pulse).

The calculation of the reflectivity $R$, transmissivity $T$ and the absorbance *1-R-T* are derived via the employment of the multiple reflection theory [23]. More specifically, the following expressions are employed to evaluate the optical parameters for a thin film placed on a thick substrate assuming a *p*-polarised beam although other polarization states can also be included (at room temperature [23, 24])

$$R = |\,r_{dl}|^2, \quad T = |t_{dl}|^2 \tilde{N}_S, \quad r_{dl} = \frac{r_{am} + r_{mS} e^{2\beta j}}{1 + r_{am} r_{mS} e^{2\beta j}},$$
$$t_{dl} = \frac{t_{am} t_{mS} e^{\beta j}}{1 + r_{am} r_{mS} e^{2\beta j}}, \quad \beta = \frac{2\pi d \tilde{N}_m}{\lambda_L} \qquad (1)$$
$$r_{am} = \frac{\tilde{N}_m - \tilde{N}_a}{\tilde{N}_m + \tilde{N}_a}, \; r_{mS} = \frac{\tilde{N}_S - \tilde{N}_m}{\tilde{N}_S + \tilde{N}_m}, \quad t_{am} = \frac{2\tilde{N}_a}{\tilde{N}_m + \tilde{N}_a}, t_{ms} = \frac{2\tilde{N}_m}{\tilde{N}_S + \tilde{N}_m}$$

where the indices '$a$', '$m$', '$S$' stand for 'air', '$SiO_2$', 'Si', respectively. The refractive indices of the materials such as air, Si and $SiO_2$ are equal to $\tilde{N}_a = 1, \tilde{N}_S$, $\tilde{N}_m$, respectively at the laser wavelength $\lambda_L$. Following the precise evaluation of the portion of the energy absorbed from the $SiO_2$/Si complex, the next component of the theoretical model is related to the description of ultrafast phenomena in the $SiO_2$ and Si materials.

#### a. Ultrafast dynamics in SiO2

To describe the free electron generation in $SiO_2$ upon excitation with ultrashort Mid-IR pulsed lasers a standard model of two rate-equations (Eq.2) is used; these single rate-equations yield both the evolution of the excited electron and self-trapped exciton (STE) densities, $N_e$ and $N_{STE}$ [25, 26], respectively,

$$\begin{aligned} \frac{dN_e}{dt} &= \frac{N_V-N_e}{N_V}\left(W_{PI}^{(1)} + N_e A^{(1)}\right) + \frac{N_{STE}}{N_V}\left(W_{PI}^{(2)} + N_e A^{(2)}\right) - \frac{N_e}{\tau_{tr}} \\ \frac{dN_{STE}}{dt} &= \frac{N_e}{\tau_{tr}} - \frac{N_{STE}}{N_V}\left(W_{PI}^{(2)} + N_e A^{(2)}\right) \end{aligned} \tag{2}$$

where $N_V$=2.2×10$^{22}$ cm$^{-3}$ stands for the valence electron density. In Eqs.2, the STE states are considered to be centres situated at an energy level below the conduction band (i.e. $E_G^{(2)}$=6 eV); it is recalled that, for $SiO_2$, the band gap between the valence (VB) and the conduction band (CB) is $E_G^{(1)}$=9 eV [25, 27] while $\tau_r$ ~150 fs [28] stands for the trapping time of electrons in STE states. In the above framework, photoexcitation assumes photoionization ($W_{PI}^{(i)}$, which can be due to multiphoton, tunneling or multiphoton/tunneling ionization) and impact ionization processes (i.e. $A^{(i)}$ stands for the avalanche ionisation rate) which can allow a transition from VB to CB ($i$=1) and from STE level to CB ($i$=2). As shown in previous reports [28-30], the electron excitation mechanisms are dependent on the laser intensity $I$. In the present model, an attenuation of the local laser intensity due to the photoionisation and inverse bremsstrahlung (Free Carrier) absorption along the depth $z$ of the dielectric is considered and described by Eq.3

$$\begin{aligned} \frac{dI}{dz} &= N_{ph}^{(1)}\hbar\omega_L \frac{N_V-N_e}{N_V} W_{PI}^{(1)} + N_{ph}^{(2)}\hbar\omega_L \frac{N_{STE}}{N_V} W_{PI}^{(2)} - \alpha(N_e)I \\ I &= \left(1 - R(t) - T(t)\right)\frac{2\sqrt{ln2}}{\sqrt{\pi}\tau_p} F e^{-4ln2\left(\frac{t-3\tau_p}{\tau_p}\right)^2} \end{aligned} \tag{3}$$

where $N_{ph}^{(i)}$ stands for the minimum number of photons required to be absorbed by an electron located in the valence band ($i$=1) or the band where the STE states reside ($i$=2) to overcome the relevant energy gap and reach the conduction band. For the sake of simplicity, films of thicknesses which are remarkably smaller than the size of the spot radius of the laser beam are considered that allows to model the multiscale processes in one dimension (along the $z$-axis). On the other hand, $F$ and $\omega_L$ correspond to the laser fluence and frequency. Finally, the last term in the first equation in Eq.3 corresponds to inverse bremsstrahlung while the carrier density dependent parameter $\alpha$ corresponds to the free carrier absorption coefficient. It is noted that the parameters $R$ and $T$ are time dependent variables as carrier excitation influences their evolution.

#### b. Ultrafast dynamics in Si

On the other hand, the ultrafast dynamics in Si is described by the following set of equations (Eq.4)

$$\begin{aligned} \frac{dN_e^{(Si)}}{dt} &= \frac{\beta_{TPA}}{2\hbar\omega_L} I^2 + \frac{\gamma_{TPA}}{3\hbar\omega_L} I^3 - \gamma\left(N_e^{(Si)}\right)^3 + \theta N_e^{(Si)} - \vec{\nabla}\cdot\vec{J} \\ \frac{dI}{dz} &= -\alpha_{FCA} I - \beta_{TPA} I^2 - \gamma_{TPA} I^3 \\ I &= T(t)\frac{2\sqrt{ln2}}{\sqrt{\pi}\tau_p} F e^{-4ln2\left(\frac{t-3\tau_p}{\tau_p}\right)^2} \end{aligned} \tag{4}$$

In the above expressions, $\beta_{TPA}, \gamma_{TPA}, \alpha_{FCA}$ stand for the two-photon, three-photon and free carrier absorption, respectively, $\vec{J}$ is the carrier current density (see below) and $T$ corresponds to the transmissivity that is also related to the absorbed part of the laser intensity from the substrate. Furthermore, $N_e^{(Si)}$ correspond to the carrier density inside Si while $\theta$ stands for the impact ionisation coefficient. For a more detailed description of the ultrafast dynamics for Si, see Ref. [9]. It is noted that, in this work, the multiphoton ionisation process considers only two-photon and three-photon absorption processes as the focus was centred on laser wavelengths between 2.2 μm and 3.2 μm, where two-photon and three-photon excitation dominates the multiphoton excitation mechanism [9, 31-34]. An extension of the model at longer wavelengths (i.e. smaller photon energies) requires the knowledge of higher order absorption coefficient.

It is known that the optical parameters of a material vary with the excitation level reached during the irradiation [35, 36] which implies that a precise evaluation of the ultrafast phenomena requires the calculation of the dynamics of reflectivity, transmissivity and absorptivity. The calculation of the optical parameters is performed through the evaluation of the dielectric function $\varepsilon$, which is related to the refractive index $n$ and extinction coefficient $k$ through the expression $\varepsilon = (n+ik)^2$. It is recalled that the refractive indices $n$ of *unexcited* Si and $SiO_2$ in Mid-IR regime are provided by the following expressions [9, 31-34, 37, 38] (Eq.5)

$$\begin{aligned} n^2 &= 11.67316 + \frac{1}{{\lambda_L}^2} + \frac{0.004482633}{{\lambda_L}^2 - 1.108205^2} && (\text{For Silicon}, 22\ \mu\text{m} \geq \lambda_L \geq 2.5\ \mu\text{m}) \\ n^2 &= 1 + \frac{10.6684293{\lambda_L}^2}{{\lambda_L}^2 - 0.301516485^2} + \frac{0.0030434748{\lambda_L}^2}{{\lambda_L}^2 - 1.13475115^2} + \frac{1.54133408{\lambda_L}^2}{{\lambda_L}^2 - 1104^2} && (\text{For Silicon}, 2.5\mu\text{m} \geq \lambda_L \geq 1.36\ \mu\text{m}) \\ n^2 &= 1 + \frac{0.6961663{\lambda_L}^2}{{\lambda_L}^2 - 0.0684043^2} + \frac{0.4079426{\lambda_L}^2}{{\lambda_L}^2 - 0.11624142^2} + \frac{0.8974794{\lambda_L}^2}{{\lambda_L}^2 - 9.896161^2} && (\text{For Fused Silica}) \end{aligned} \tag{5}$$

On the other hand, the expression that provides the dynamics of the dielectric function due to the variation of the density of the carriers in an excited material $N_e^{(a)}$ is given by Eq.6 ($a$=1 for $SiO_2$ and $a$=2 for Si)

$$\varepsilon^{(a)} = 1 + \left(\varepsilon_{un}^{(a)} - 1\right)\left(1 - \frac{N_e^{(a)}}{N_V^{(a)}}\right) \frac{e^2 N_e^{(a)}}{m_r m_e \varepsilon_0 {\omega_L}^2} \frac{1}{\left(1 + i\frac{1}{\omega_L \tau_c}\right)} \tag{6}$$

In Eq.6, $\varepsilon_{un}^{(a)}$ corresponds to the dielectric parameter of the unexcited material that is $\lambda_L$-dependent which is calculated from Eq.5, $N_e^{(a)}$ and $N_V^{(a)}$ are the carrier densities in the conduction and valence bands ( $N_V^{(1)} = 2.2 \times 10^{22}\text{cm}^{-3}$ [25] and $N_V^{(2)} = 5 \times 10^{22}\text{cm}^{-3}$ [36], respectively, $m_e$ is the electron mass, $e$ is the electron charge, $m_r = 0.18$, $\varepsilon_0$ is the vacuum permittivity, $\omega_L$ is the laser frequency and $\tau_c = 1.1$ fs stands for the electron collision time. It is recalled that various values for $\tau_c$ have been used in previous reports (see Ref. [26] and references therein). Furthermore, it is noted that the contribution of Kerr effect for Mid-IR fs pulses in the dielectric function has been taken into account for Si [9] and $SiO_2$ [26].

Finally, the thermal response of the $SiO_2$/Si is described via the employment of the traditional Two-Temperature Model (TTM) for the two materials. In particularly, the relaxation-time approximation to Boltzmann's transport equation is used to provide the electron $T_e^{(Si)}$ and lattice $T_L^{(Si)}$ temperature, respectively, of Si

$$\begin{aligned} C_e^{(Si)} \frac{\partial T_e^{(Si)}}{\partial t} &= \frac{\partial}{\partial z}\left(k_e^{(Si)} \frac{\partial}{\partial z} T_e^{(Si)}\right) - G_{cL}\left(T_e^{(Si)} - T_L^{(Si)}\right) + S \\ C_L^{(Si)} \frac{\partial T_L^{(Si)}}{\partial t} &= \frac{\partial}{\partial z}\left(k_L^{(Si)} \frac{\partial}{\partial z} T_L^{(Si)}\right) + G_{cL}\left(T_e^{(Si)} - T_L^{(Si)}\right) \end{aligned} \tag{7}$$

where $C_e^{(Si)}$ and $C_L^{(Si)}$ stand for the heat capacities of the carriers and the lattice, respectively, $G_{cL}$is the carrier-phonon coupling, $k_L^{(Si)}$, ($k_e^{(Si)}$) correspond to the lattice's (carrier's) thermal conductivity and $S$ stands for a generalized 'source' term given by the following expression [9]

$$\begin{aligned} S &= \alpha_{FCA} I + \beta_{TPA} I^2 + \gamma_{TPA} I^3 - \gamma\left(N_e^{(Si)}\right)^3 - \vec{\nabla} \cdot \vec{W} - \frac{\partial N_e^{(Si)}}{\partial t}\left(E_g + 3k_B T_e^{(Si)}\right) - N_e^{(Si)}\left(\frac{\partial E_g}{\partial T_L^{(Si)}} \frac{\partial T_L^{(Si)}}{\partial t} + \frac{\partial E_g}{\partial N_e^{(Si)}} \frac{\partial N_e^{(Si)}}{\partial t}\right) \\ \vec{W} &= \left(E_g + 4k_B T_e^{(Si)}\right)\vec{J} - \left(k_e^{(Si)} + k_h^{(Si)}\right)\vec{\nabla} T_e^{(Si)} \\ \vec{J} &= D\left(\vec{\nabla} N_e^{(Si)} + \frac{N_e^{(Si)}}{2k_B T_e^{(Si)}} \vec{\nabla} E_g + \frac{N_e^{(Si)}}{2T_e^{(Si)}} \vec{\nabla} T_e^{(Si)}\right) \end{aligned} \tag{8}$$

where $E_g$ stands for the energy gap of Silicon, $k_B$ is the Boltzmann constant, $D$ is the ambipolar carrier diffusivity.

By contrast, a similar TTM which describes the electron and lattice temperature evolution for fused silica [26] is not used in this work. This is due to the fact that simulation predictions demonstrate that no energy absorption (i.e. following the investigation of the dynamics of the reflectivity and transmissivity via Eqs.1-6; simulations always yield $R(t)+T(t)=1$ for all film thicknesses, see, also simulation results in the Supplementary Material) occurs in the dielectric material which implies that no electron excitation is produced. Therefore, no noticeable thermal effects are expected either in the electron or the lattice system and the only but insignificant increase in the latticed temperature of $SiO_2$ takes place near the interface due to heat transfer from the substrate (see next Section).

## III. SIMULATION PROCEDURE

To solve Eqs.1-8, an iterative Crank-Nicolson scheme is used which is based on a finite-difference method [24]. A small and insignificant variation of the lattice temperature on $SiO_2$, mostly, at the back end of the film is calculated through the boundary conditions considered on the interface between the top layer and the substrate: $k_L \frac{\partial T_L}{\partial z} = k_L^{(Si)} \frac{\partial T_L^{(Si)}}{\partial z}$, where $k_L$ stands for the lattice heat conductivity in $SiO_2$, respectively, while $T_L$ corresponds to the lattice temperature in $SiO_2$. Finally, it is assumed that there is no carrier diffusion from $SiO_2$ into Si. Thus, the boundary condition $\frac{\partial N_e^{(Si)}}{\partial z} = \frac{\partial N_e}{\partial z}$ =0 is considered on the interface. Due to the absence of electron excitation in fused silica at all times this boundary condition turns to: $\frac{\partial N_e^{(Si)}}{\partial z}$ =0 (and $N_e = 0$). It is noted that in this work simulations were performed, mainly, assuming laser pulses of wavelength $\lambda_L$=3.2 μm (unless otherwise stated and pulse duration $\tau_p$=170 fs).

## IV. RESULTS

Simulations are firstly performed at $\lambda_L = 3.2$ μm where the refractive indices of the materials such as air, Si and $SiO_2$ are equal to $\tilde{N}_a = 1, \tilde{N}_S = 3.4309$ [39], $\tilde{N}_m = 1.4143$ [40], respectively (at room temperature). Although the predominant focus of the current work is centred on the investigation of the response of the $SiO_2$/Si at $\lambda_L$=3.2 μm, a similar exploration can be performed at other laser wavelengths. Without a loss of generality calculations are performed for $d$ between 20 nm and 5 μm. Simulation results predict some interesting features of the optical parameters which are illustrated in Fig.2. More specifically, results derived using Eq.1 for $R$ and $T$ at 300 K (before the material is irradiated) indicate that $R+T$=1 (Fig.2a); these results confirm that no energy is absorbed from $SiO_2$ (at 300 K). According to the theoretical predictions (Fig.2a), the range of values of the reflectivity lie between 0.07 and 0.3 while the amount of the energy transmitted into the substrate varies between 0.7 and 0.93 of the deposited energy. These calculations manifest that the antireflection properties of the complex can be controlled by an appropriately selected thickness of the $SiO_2$ film. Interestingly, there exists a pronounced periodicity of the optical parameter dependence as a function of thickness demonstrated in Figs.2a,b,d; this is due to the term $\beta = \frac{2\pi d \tilde{N}_m}{\lambda_L}$ in the expression that provides the reflectivity and the transmissivity (Eq.1). According to Eq.1, $\beta$ is included in $e^{2\beta j}$ which leads to a (spatially) periodic behaviour of the optical parameters. Thus, the part of the laser energy (at room temperature) is expected to follow a similar trend and, in the next paragraphs, it will be shown that this periodicity will also be projected on the excitation (i.e. carriers) and thermal response (including the damage threshold) of the system. Theoretical results depicted in Fig.2b,c,d indicate that the periodicity at increasing wavelength increases. This is due to the fact that $\widetilde{N}_m$ drops at increasing $\lambda_L$ which yields a smaller value of $\beta$ and therefore, a larger periodicity. More specifically, an analysis of the predictions at $d$=3 μm (*white* dashed line in Fig.2b and produced results shown in Fig.2c) confirms the dependence of the reflectivity as increasing wavelength for the same $SiO_2$ film thickness. A similar behaviour was also reported in a previous work for a different configuration where a Si film of thickness $d$ was placed on top of a $SiO_2$ substrate [17].

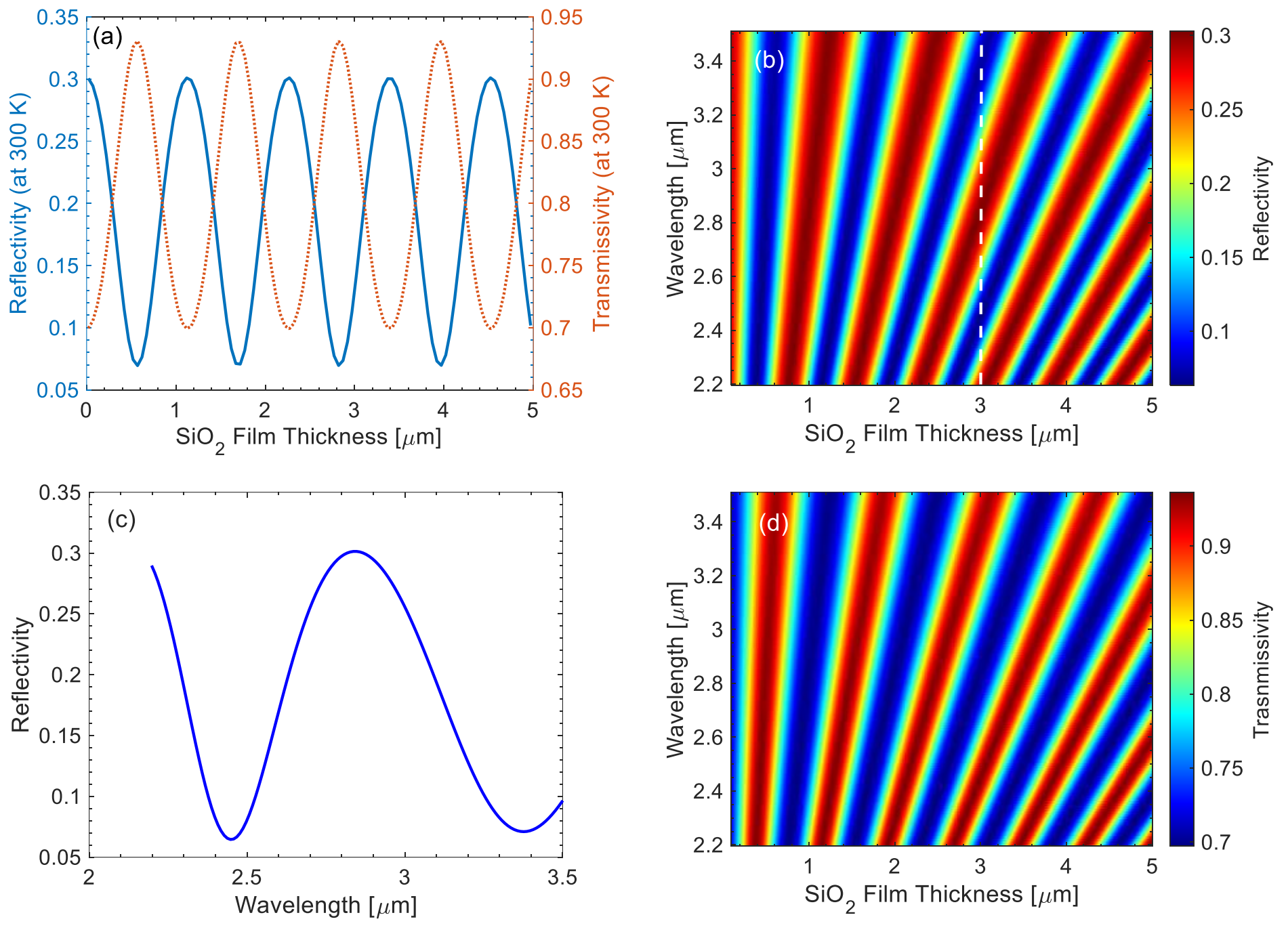

Figure 2: (a) Optical parameters of $SiO_2$/Si as a function of $SiO_2$ thickness (for $\lambda_L$=3.2 μm), (b) Reflectivity of $SiO_2$/Si as a function of $d$ and the laser wavelength, (c) Reflectivity of $SiO_2$/Si at various wavelengths along the *white* dashed line in (b), at $d$=3 μm), (d) Transmissivity of $SiO_2$/Si as a function of $d$ and the laser wavelength. All calculations are at room temperature.

To explore the damage conditions following irradiation of the two-layered complex with Mid-IR fs pulses, it is important to analyse the ultrafast phenomena and derive the induced thermal response of the system. Results in Fig.3 provide the dynamics of the electron density on the surface of the semiconducting material (Fig.3a) and the temporal evolution of $N_e^{(Si)}$ *inside* Si (Fig.3b) assuming irradiation of $SiO_2$/Si with fluence $F$=0.1 J/cm$^2$ considering a $SiO_2$ film of $d$=520 nm. Similar results are deduced (results are not shown here) at other $d$ and $F$ values for which a $d$-dependent maximum electron density is produced as the absorbed energy from $SiO_2$ varies with $d$. It is evident that unlike fused silica, Si absorbs significantly leading to high levels of excitation (~1.05×10$^{21}$ cm$^{-3}$). The spatiotemporal carrier density distribution inside 1μm of Si within 3 ps is illustrated in Fig.3b. Similarly, the lattice temperature distribution is depicted for the same laser conditions in Fig.3c.

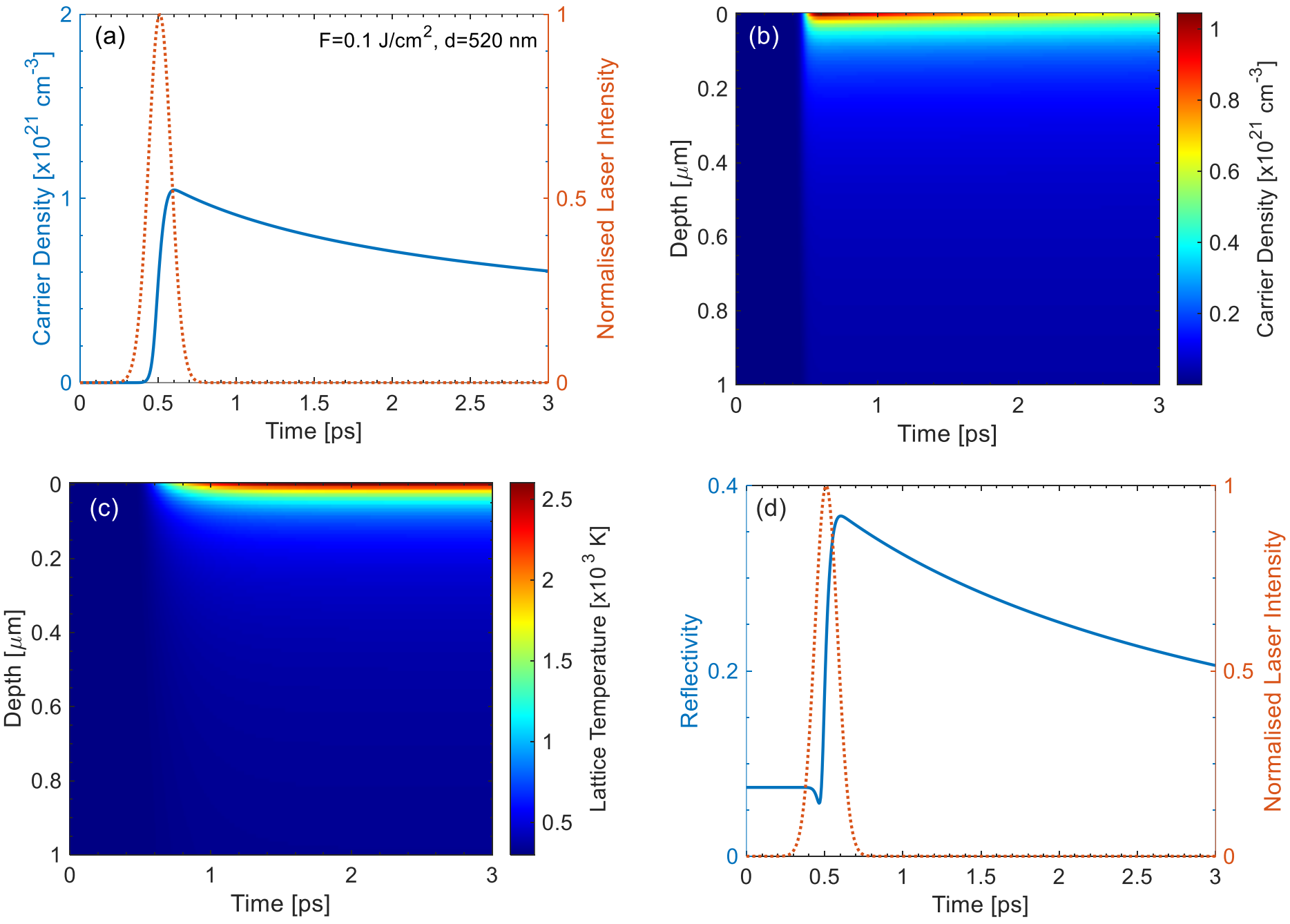

Figure 3: (a) Electron density evolution on the surface of Si. Electron density (b) and Lattice temperature (c) evolution inside Si. (d) Reflectivity dynamics of $SiO_2$/Si. ($F$=0.1 J/cm$^2$, $d$=520 nm).

A key issue that requires a particular investigation is the trend of the optical parameters *during* the pulse. It is noted that results shown in Fig.2 focused on an analysis of the values of $R$ and $T$ at room temperature. Nevertheless, a more precise exploration should analyse the fingerprint of the excitation of carriers and ultrafast dynamics on the temporal variation of $R$ and $T$. With respect to the irradiation of semiconducting and dielectric targets with ultrashort laser pulses, it has been demonstrated [9, 35, 36] that the excitation of carriers yields a variation of the dielectric function (Eq.6) and therefore the optical parameters which influence, eventually, the energy absorption from the material (i.e. $\widetilde{N}_S$ in Eq.1 becomes a complex number). More specifically, assuming irradiation of the two-layered stack with $F$=0.1 J/cm$^2$ considering a $SiO_2$ of $d$=520 nm, simulation results show there is a significant variation of the reflectivity of the complex shown in Fig.3d. The temporal analysis of the optical parameters following the detailed investigation of the ultrafast dynamics show that during the pulse duration the sum of reflectivity and the transmissivity is equal to 1 that demonstrates there is no absorption of energy from $SiO_2$ and therefore $N_e$ is always zero (see Supplementary Material).

Observing the remarkable variation in the reflectivity during the pulse duration stated in the previous paragraph and before analysing the correlation of the thermal response and the minimum energy required to damage Si with the coating thickness via the employment of fs pulses, it is important to evaluate the levels of the reflected/transmitted amounts of the laser energy derived from the density of excited electrons values. In Fig.4, the minimum and maximum reflectivity values are depicted for $d$ between 20 nm and 5 μm for $F$=0.06 J/cm$^2$, 0.08 J/cm$^2$, 0.1 J/cm$^2$ (at the end of the section, a discussion is presented on the selection of the tree fluence values).

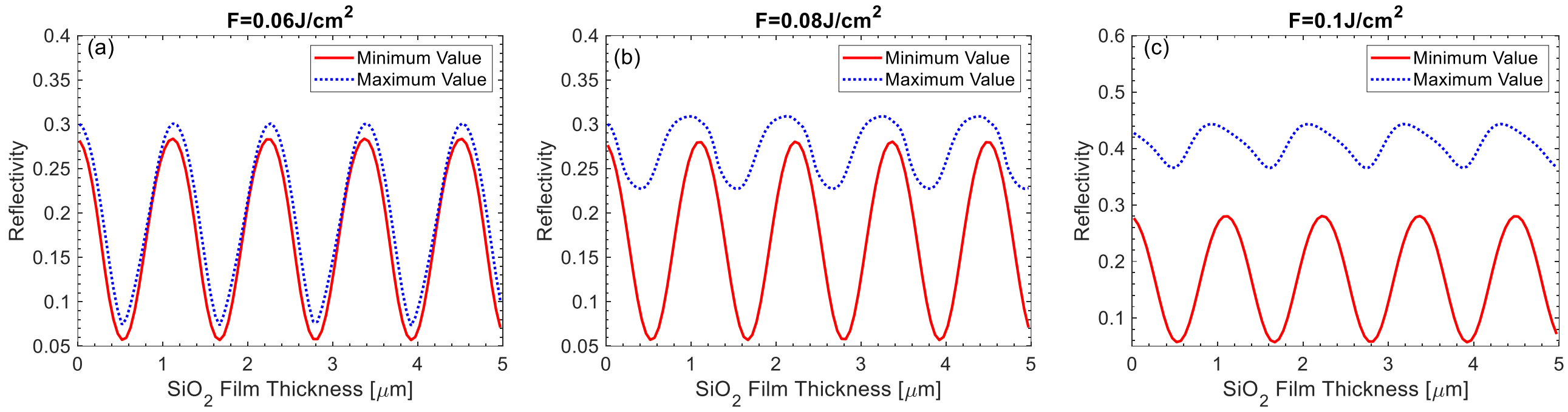

Figure 4: Maximum and Minimum values of Reflectivity of Si/ $SiO_2$ at various fluences ((a) $F$=0.06 J/cm$^2$, (b) $F$=0.08 J/cm$^2$, (c) $F$=0.1 J/cm$^2$) as a function of the $SiO_2$ thickness. Simulations are illustrated at $\lambda_L$=3.2 μm.

Results show that the reflectivity temporal variation at the lowest fluence is negligible while the increasing excitation at higher fluences leads to an enhanced reflectivity change. Thus, at $F$=0.06 J/cm$^2$, there is a rather insignificant variation of $R$ for some $d$ while there is a more pronounced change at other $SiO_2$ thicknesses.

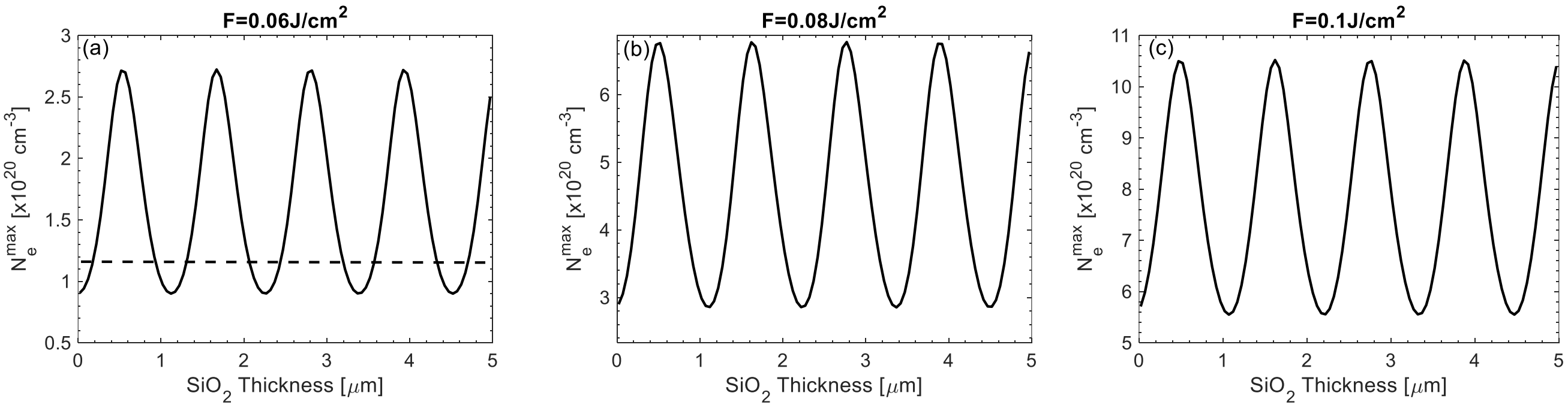

Figure 5: Maximum Carrier density on the Si surface at various fluences ((a) $F$=0.06 J/cm$^2$, (b) $F$=0.08 J/cm$^2$, (c) $F$=0.1 J/cm$^2$) as a function of the $SiO_2$ thickness. Simulations are illustrated at $\lambda_L$=3.2 μm. *Dashed* line in (a) indicates OBT.

In addition to the optical parameter variation at different fluences and coating thicknesses, the dependence of the electron excitation and thermal response of the system on $d$ are also evaluated. Simulation results illustrated in Fig.5 show a periodic variation of the carrier density as a function of $d$ for all fluences considered in this work. This is due to the interference effects expressed in the optical parameters (Eq.1) and, therefore, energy absorption $d$-dependence. The values of the maximum values of $N_e^{(Si)}$ (denoted as $N_e^{max}$) range between ~10$^{20}$ cm$^{-3}$ and ~1.5×10$^{21}$ cm$^{-3}$. The black *dashed* line in Fig.5a corresponds to the critical value $N_e^{cr}$(i. e. $N_e^{cr} \equiv \frac{4\pi^2 c^2 m_e \varepsilon_0}{(\lambda_L^2 e^2)}$ which is, usually, termed as the optical breakdown threshold (OBT) [28]. In the expression that provides $N_e^{cr}$, $c$ is the speed of light, $m_e$ stands for mass of electron, $e$ is the electron charge and $\varepsilon_0$ is the permittivity of vacuum. In particularly, $N_e^{cr} = 1.09 \times 10^{20}$ cm$^{-3}$ at $\lambda_L = 3.2$ μm. In previous reports, and more specifically, for dielectrics [30, 41-44], the OBT has been associated with the minimum critical density that leads to material damage and it was, therefore, linked with DT of transparent materials. By contrast, in other works [25, 27, 45, 46], a more precise evaluation of DT was derived via a thermal criterion (i.e. fluence at which the material starts to melt). To evaluate whether the two criteria are similar, the thermal response of the system is calculated. Results at the same fluences as before ($F$=0.06 J/cm$^2$, $F$=0.08 J/cm$^2$ and $F$=0.1 J/cm$^2$ see Fig.6a,b,c), demonstrate, again, a periodic dependence of the maximum lattice temperature $T_L^{(Si)}$ (denoted as $T_L^{max}$), however, the melting point (associated directly with DT) is reached in different conditions (black dotted line Fig.6b,c). Thus, a comparison of Figs.5,6 show that the OBT-based threshold provides an underestimation of the DT. According to Fig.6b, $F$=0.08 J/cm$^2$ represents a fluence value at which the material starts to melt at particular values of $d$ characterized by a periodicity.

To evaluate the influence of the coating on both the optical response of the complex and the thermal effects on the substrate, simulation results are illustrated in Fig.6d for two cases, where the evolution of the lattice temperature and transmissivity in the Si material are shown: (a) assuming the presence of a $SiO_2$ film of thickness $d$=520 nm and (b) in absence of the fused silica coating. Interestingly, results indicate that for this particular $SiO_2$ thickness, the transmissivity increases leading to a larger energy absorption from the substrate which results to a larger maximum temperature. Similar conclusions can be deduced for other values of *d*. Results for the reflectivity (see Supplementary Material) demonstrate that *R(t)*+*T(t)*=1. The above results show that the selection of the fluence values was based on the distinct thermal response for each case. More specifically, at *F*=0.06 J/cm$^2$, the maximum lattice temperature which is attained is not sufficient to melt the material. On the other hand, at *F*=0.08 J/cm$^2$ melting occurs for a small range of values of the coating thickness *d* (Fig.6b). Finally, at *F*=0.1 J/cm$^2$ melting rises independently of *d*.

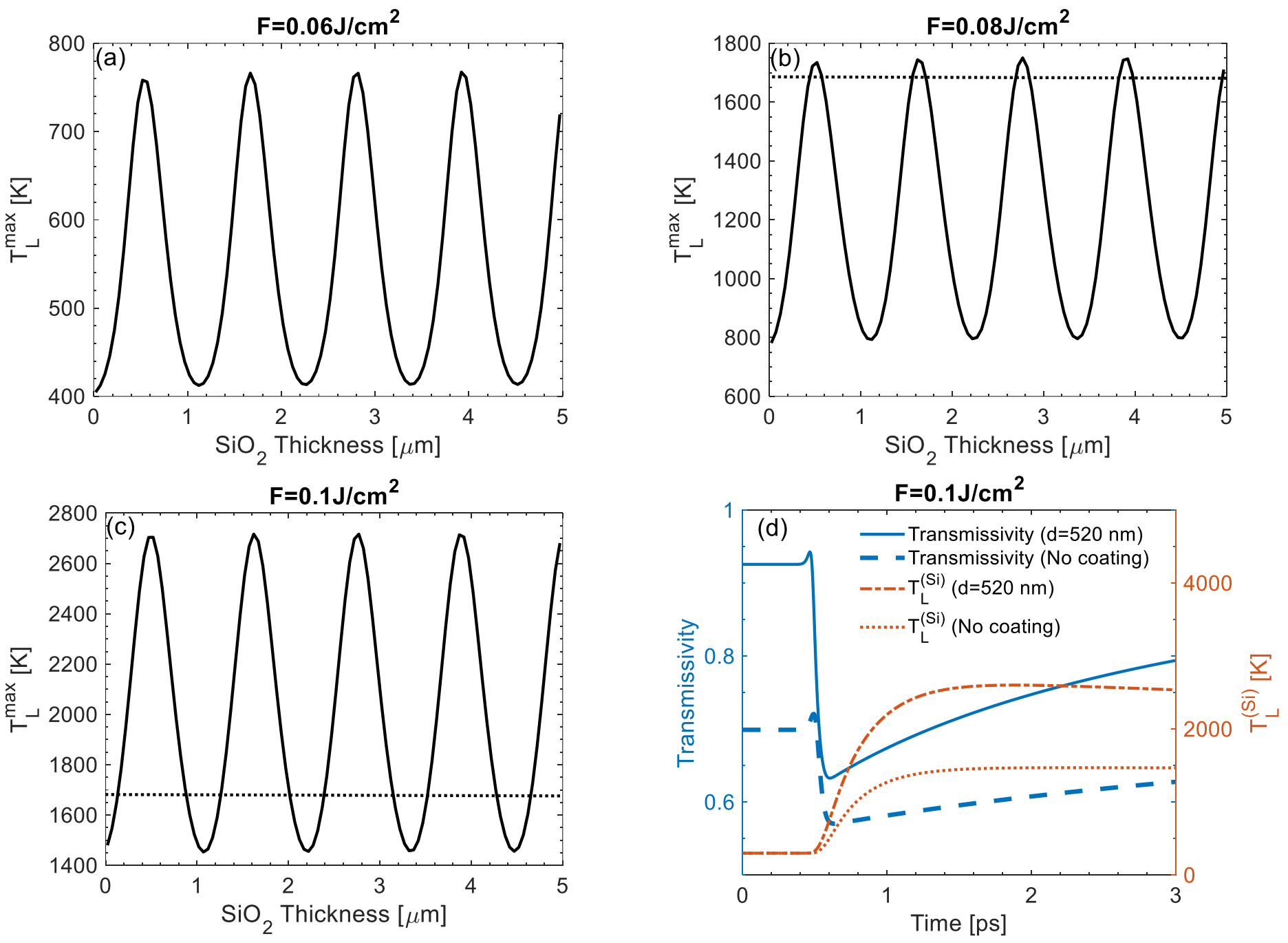

Figure 6: Maximum lattice temperature on the Si surface at various fluences ((a) *F*=0.06 J/cm$^2$, (b) *F*=0.08 J/cm$^2$, (c) *F*=0.1 J/cm$^2$) as a function of the $SiO_2$ thickness. *Dotted* line in (b,c) indicates melting point. (d) Transmissivity in Si for *d*=520 nm and in absence of $SiO_2$ (*F*=0.1 J/cm$^2$). Simulations are illustrated at $\lambda_L$=3.2 μm.

While the aforementioned analysis was performed in laser conditions assuming the same photon energy and pulse duration, an evaluation of the ultrafast dynamics and thermal response of the system at different laser wavelengths and pulse duration are expected allow the determination of the impact of those parameters on the physical processes.

### i. Dependence of $N_e^{max}$, $T_L^{max}$ and DT on the pulse duration

In Fig.7a, the maximum density of the excited carriers on Si, $N_e^{max}$, as a function of the coating thickness at three different values of the pulse duration (i.e. $\tau_p$=170 fs, 500 fs, 1 ps) at $\lambda_L = 3.2$ μm and at *F*=0.1 J/cm$^2$ are illustrated. The three pulse durations were appropriately selected to illustrate the role of short, long and intermediate pulses in the ultrafast dynamics. Due to the fact that longer pulse durations yield smaller laser intensity and, eventually, energy absorption from Si, the maximum carrier density is expected to drop at increasing $\tau_p$ for the same *d*. A similar trend is revealed for the maximum lattice temperature $T_L^{max}$ (Fig.7b). As expected, both parameters exhibit a periodic dependence on the $SiO_2$ thickness. Finally, for the three pulse durations, DT as a function of *d* is illustrated in Fig.7c. The horizontal *dotted* lines correspond to the DT of Si, $F_{\tau_p}^{(Si)}$, following irradiation with $\tau_p$=170 fs, 500 fs, 1 ps, respectively, in *absence* of the $SiO_2$ coating (i.e. $F_{\tau_p=170\ \mathrm{fs}}^{(Si)}$ =0.11 J/cm$^2$, $F_{\tau_p=500\ \mathrm{fs}}^{(Si)}$~0.18 J/cm$^2$, $F_{\tau_p=1\ \mathrm{ps}}^{(Si)}$ ~0.25 J/cm$^2$). As expected, $F_{\tau_p}^{(Si)}$ exhibit an increasing trend at increasing pulse temporal width. Simulation results in Fig.7c manifest a noticeable variation of DT with *d*. For all three values of the pulse duration considered in this work, in principle, there is a *decrease* of DT with a respect to the case in which the fused silica coating is absent. An analysis of the results illustrated in Fig.7c show that a significant drop of DT by a maximum 26% (for $\tau_p$=170 fs), 27% (for $\tau_p$=500 fs), 29% (for $\tau_p$=1 ps) can be achieved if coating is placed on the substrate. Furthermore, the variation of DT with the coating thickness for the three pulse durations emphasises the role of the distinct ultrafast phenomena at different excitation conditions and *d*. Furthermore, unlike predictions reported in a previous work (i.e. for a Si/$SiO_2$ [17]), the presence of the coating leads to a *drop* of the threshold for inducing damage on the substrate. This can be attributed to the different optical properties of the complex due to the placement of a lower refractive index material on the Si substrate which leads to an enhanced energy absorption from the substrate (Fig.6d).

Interestingly, for all three pulse durations in this investigation, the maximum carrier density produced on Si is higher than the OBT, $N_e^{cr}$. More specifically, $N_e^{max}$ for $\tau_p$=170 fs, 500 fs, 1 ps are calculated to be equal to $N_e^{max} \sim 6.5 \times 10^{20}$ cm$^{-3}$, $3.8 \times 10^{20}$ cm$^{-3}$, $2.8 \times 10^{20}$ cm$^{-3}$, respectively. These results serve as a justification that the determination of DT via a thermal criterion constitutes a more accurate methodology than the OBT; results, also, confirm that the OBT-based evaluation of the DT presents a necessary but not sufficient condition for the onset of damage.

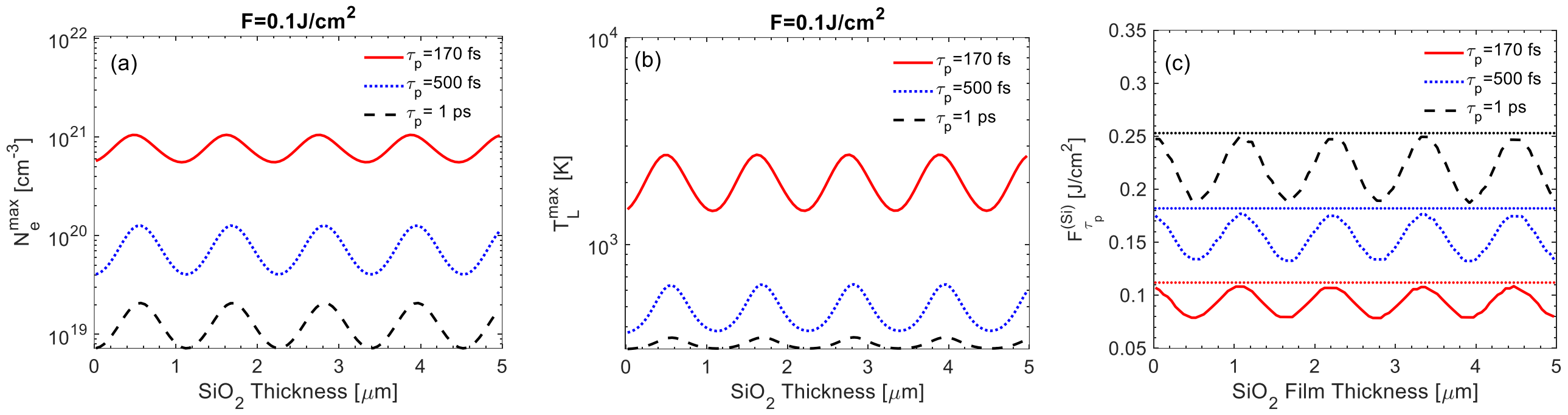

Figure 7: Maximum Carrier density (a) and lattice temperature (b) on the Si surface (*F*=0.1 J/cm$^2$) and (c) damage threshold calculation as a function of the $SiO_2$ thickness at various pulse durations ($\tau_p$=170 fs, 500 fs, 1 ps). Horizontal *dotted* lines above the curves in (c) correspond to DT of Si in absence of the coating for corresponding $\tau_p$. Simulations are illustrated at $\lambda_L$=3.2 μm.

### ii. Dependence of $N_e^{max}$, $T_L^{max}$ and DT on the laser wavelength

In the previous sections, the ultrafast dynamics and the DT on Si following irradiation of a $SiO_2$/Si complex at $\lambda_L$ =3.2 μm was systematically analysed. To evaluate the influence of the laser photon energy on the ultrafast dynamics and thermal response of the two-layered complex, a detailed analysis of the maximum density of the excited carriers on Si as a function of the coating thickness at three different laser wavelengths was performed (i.e. $\lambda_L = 2.2$ μm, 2.6 μm, 3.2 μm) at $\tau_p$=170 fs and at *F*=0.1 J/cm$^2$. These three laser wavelengths were selected to investigate the response of the system at various photon energies with distinctly different characteristics of absorption rates: for $\lambda_L = 2.2$ μm, two- and three-photon absorption coefficients are both nonzero, for $\lambda_L = 2.6$ μm only the three photon absorption coefficient is not vanishing while for $\lambda_L = 3.2$ μm a significantly smaller three-photon absorption coefficient in comparison to the previous case is nonzero [9]. Furthermore, laser sources with wavelengths in the range $\lambda_L = 2.2 - 3.2$ μm of an appreciable intensity are currently available and therefore, the simulation predictions can be tested.

Before discussing the predictions of the DT on Si at the three wavelengths, it is recalled that the theoretical framework accounting for ultrafast dynamics and thermal response for Si (Eqs.6-8) (in the absence of a $SiO_2$ coating) described adequately the damage of Si at Mid-IR [9, 12]. This demonstrates that the model which has been used successfully to describe ultrafast phenomena at lower wavelengths [6, 7] can still be employed to describe physical phenomena at smaller photon energies.

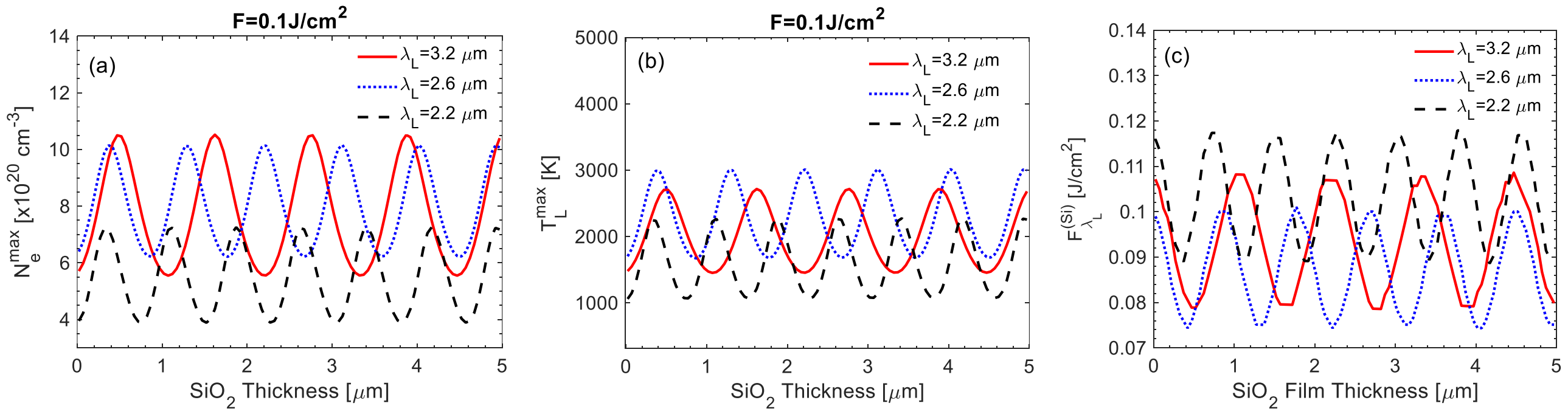

Figure 8: Maximum Carrier density (a) and maximum Carrier lattice temperature (b) on the Si surface (*F*=0.1 J/cm$^2$) and (c) damage threshold as a function of the $SiO_2$ thickness at various laser wavelength ($\lambda_L$=2.2 μm, 2.6 μm, 3.2 μm). Simulations are illustrated at $\tau_p$=170 fs.

Results for the maximum carrier density and lattice temperatures attained and the induced DT at various laser wavelengths and coating thicknesses are illustrated in Fig.8a. As explained above, the increase of the periodicity of the optical parameters demonstrated at increasing laser wavelength (Fig.2b,c,d) is also exhibited in the absorbed energy, the carrier density (Fig.8a) and lattice temperature (Fig.8b) and the damage threshold (Fig.8c).

To elaborate further on the dependence of the magnitude of the $N_e^{max}$, $T_L^{max}$ on the coating thickness at various laser wavelengths, it is important to correlate the dominant processes and the magnitude of the absorbed energy from the substrate. The remarkable smaller $N_e^{max}$, $T_L^{max}$ at the shortest laser wavelength ($\lambda_L = 2.2$ μm) for the same $d$ might be confusing as a larger photon energy is possibly expected to facilitate the electron excitation leading to a smaller DT. A conclusive interpretation of the results and the behaviour of $N_e^{max}$, $T_L^{max}$ with respect to the laser wavelength and $d$ requires a detailed analysis of: (a) the energy absorbed from the substrate for the same $SiO_2$ thickness at different wavelength and (b) the evaluation of other processes which can be more efficient in the production of excited carriers than the multiphoton ionization.

To evaluate the role of the latter, simulations conducted for three representative values of the coating thickness *d*=20 nm, 720 nm and 1220 nm (similar results can be derived for other *d*) at $\lambda_L = 2.2$ μm, 2.6 μm, 3.2 μm for 0.1 J/cm$^2$ showed that the impact ionization does not account for this behavior. More specifically, an analysis of the contribution of the impact ionization process manifests that the inclusion of this ionisation mechanism simply increases the carrier density as expected; it does not alter, though, the wavelength-dependent carrier density order at any coating thickness (see Supplementary Material). By contrast, results of the transmissivity evolution for the three values of thickness and at $\lambda_L =$ 2.2 μm, 2.6 μm, 3.2 μm for 0.1 J/cm$^2$ (see Supplementary Material) show that the monotonicity of $N_e^{max}$ and evolution of $N_e^{(Si)}$ are attributed to the amount of energy absorbed from the material. Analysing the monotonicity of the carrier density curves for the three representative (but of distinct interest due optical parameter behaviour) thicknesses, it is noted that:

(i) For *d*=20 nm, the absorbed energy from Si is *almost* identical for all three wavelengths at room temperature. When the pulse is switched on and till almost the maximum of the laser energy is reached (i.e. peak of the laser intensity), the absorbed energy will be the largest for $\lambda_L = 2.6$ μm followed by that for $\lambda_L = 3.2$ μm. On the contrary, for $\lambda_L = 3.2$ μm, the smallest amount of energy will be absorbed. This behaviour will be projected on the $N_e^{max}$ (Fig.8a); thus, $N_e^{max}$ ($\lambda_L = 2.6$ μm)> $N_e^{max}$ ($\lambda_L = 3.2$ μm) >$N_e^{max}$ ($\lambda_L =$ 2.2 μm).

(ii) For *d*=720 nm, the absorbed energy from Si is *not* identical for the three wavelengths at room temperature. More specifically, the transmissivity is larger for $\lambda_L = 3.2$ μm followed by that for $\lambda_L = 2.6$ μm and then for $\lambda_L = 2.2$ μm. When the pulse is switched on and till almost the maximum of the laser energy (i.e. peak of the laser intensity), this monotonicity is preserved resulting into the aforementioned behaviour being projected on the $N_e^{max}$ (Fig.8a).

(iii) For *d*=1220 nm, the absorbed energy from Si is *not* again identical for the three wavelengths at room temperature. The transmissivity for $\lambda_L = 2.2$ μm is slightly larger than the transmissivity for $\lambda_L = 2.6$ μm but significantly larger than that of $\lambda_L = 3.2$ μm (at room temperature). When the pulse is switched on, the first laser wavelength ($\lambda_L = 2.2$ μm) will lead to a very small increase in the absorption level compared to the case before the application of the pulse. By contrast, the second wavelength ($\lambda_L =$ 2.6 μm) which has not a very significantly different initial transmissivity from that of $\lambda_L = 2.2$ μm will lead to an increase of the absorption level (Fig.8a). Thus, the produced $N_e^{max}$ for $\lambda_L = 2.6$ μm will be higher than that of $\lambda_L = 2.2$ μm. Finally, the third wavelength ($\lambda_L = 3.2$ μm) that yields a significantly smaller transmissivity gives $N_e^{max}$ smaller than the other two wavelengths.

### iii. Excitation of SP modes

One important aspect in material processing is the fabrication of subwavelength structures via the use of laser pulses. As excitation of surface plasmons and their interference with the incident beam is regarded as the predominant mechanism of periodic structure formation [7, 47], the periodicities of the (allowed) excited electromagnetic surface modes could provide an estimate of the expected sub-wavelength structures on the irradiated material. Considering the possibility of excitation of SP modes on an air/$SiO_2$/Si stack, it is noted that such electromagnetic modes can be excited on the $SiO_2$/Si interface if a sufficiently high $N_e$ is reached to metallise the semiconductor [7, 47]. While a simplistic scenario is to treat the substrate as a material that turns (entirely or a large part of it) into a metallic solid, a more precise description is to assume that a thin layer of metallic Si is formed while the rest of the substrate maintains the properties of a semiconductor [48, 49]. Thus, it is assumed that the excitation of Si leads to the formation of: (i) a volume of Si where $Re(\varepsilon_{Si})<0$ due to a sufficient amount of excited carrier density that metallises the semiconductor and (ii) a volume of non-metallic Si (termed as $Si^{(non\text{-}met)}$) where $Re(\varepsilon_{Si})>0$. Therefore, the final configuration which should be investigated for SP excitation is an 'air/$SiO_2$/ (metallic Si)/$Si^{(non\text{-}met)}$ stack that allows the excitation of SP on both interfaces $SiO_2$/ (metallic Si) and (metallic Si)/$Si^{(non\text{-}met)}$ leading, potentially, to a coupling of the SP modes if the metallic layer thickness is not large enough [50-52] (i.e. for the sake of simplicity, hereafter, '$Si^{(non\text{-}met)}$' will be termed as 'Si'). In a previous report [48], results showed that the consideration of a thin layer of metallic Si (~10 nm) can influence the SP periodicity in contrast to the behaviour for thicker layers (~100 nm) following irradiation at 800 nm. Nevertheless, a more precise investigation is necessitated to evaluate the characteristics of the excited at longer pulses and discuss the role of thin metallic layers.

The features of the localised SP modes is calculated via the following dispersion relation Eq.(9) which was generated by assuming the continuity of the tangential components of the magnetic ($H_y$) and electric ($E_x$) fields on the two interfaces

$$e^{-2k_2 d} = \frac{\left(1+\frac{k_3/\varepsilon_3}{k_2/\varepsilon_2}\right)\left(1+\frac{k_4/\varepsilon_4}{k_3/\varepsilon_3}\right)+\left(1-\frac{k_3/\varepsilon_3}{k_2/\varepsilon_2}\right)\left(1-\frac{k_4/4}{k_3/\varepsilon_3}\right)e^{-2k_3 d_3}}{\left(1+\frac{k_4/\varepsilon_4}{k_3/\varepsilon_3}\right)\left(1-\frac{k_3/\varepsilon_3}{k_2/\varepsilon_2}\right)+\left(1-\frac{k_4/\varepsilon_4}{k_3/\varepsilon_3}\right)\left(1+\frac{k_3/\varepsilon_3}{k_2/\varepsilon_2}\right)e^{-2k_3 d_3}} \times \frac{\left(1+\frac{k_2/\varepsilon_2}{k_1/\varepsilon_1}\right)}{\left(-1+\frac{k_2/\varepsilon_2}{k_1/\varepsilon_1}\right)} \tag{9}$$

where $k_j = \sqrt{\tilde{\beta}^2 - \varepsilon_j k_0^2}$ ($j$=1, 2, 3, 4 for 'air', '$SiO_2$', 'metallic Si', and 'Si'), $\varepsilon_j$ stands for the dielectric permittivity of the $j$ material (derived from Eq.6), $d_3$ is the thickness of the layer of the 'metallic Si', $k_0$ (=$2\pi/\lambda_L$) is the free-space wavenumber at the laser wavelength and $\tilde{\beta}$ is the propagation constant of the SP. Thus, Eq.9 yields the supported SP solutions; each solution is associated with an SP wavelength provided by $\Lambda = \frac{2\pi}{Re(\tilde{\beta})}$. It is noted that only *bound* modes of the SPs are considered [52]. Eq.9 indicates that the dispersion relation of SPs can be tailored by using different thicknesses of the dielectric coating $d$. Furthermore, it shows that the properties of the SP are influenced by both the air and the dielectric depending on the thickness of the coating and the laser frequency $\omega_L$ (=$2\pi c/\lambda_L$, in air). In particularly, results presented in a previous report [20] revealed that in an 'air/dielectric/metal' stack, for low $\omega_L$ (thus, long wavelength) and small $Re(\tilde{\beta})$the excited SP mode 'sees' the air first and the dispersion curve follows that of an air/metal stack. By contrast, at higher $\omega_L$ (thus, shorter wavelength) and larger $Re(\tilde{\beta})$, the SP does not 'see' significantly the outside material while the dispersion curve asymptotically behaves as the curve for $SiO_2$/metal.

For an 'air/$SiO_2$/(metallic Si)/Si' stack, simulations have been conducted to evaluate, firstly, the minimum density of the excited carriers to alter Si, from a semiconducting to a metallic state. This critical density, $N_e^{(m-Si)}$, is derived from the Drude model (Eq.6) for Si assuming the real part of the dielectric function of the excited silicon becomes negative ($Re(\varepsilon_{Si}) < 0$). In Fig.9a, the dependence of $N_e^{(m-Si)}$ on $\lambda_L$ is illustrated and a fitting procedure manifests the following correlation

$$N_e^{(m-Si)}(\lambda_L) = 27.5\, e^{-2.706\lambda_L} + 1.597\, e^{-0.2298} \ (\times 10^{21}\ cm^{-3}) \tag{10}$$

In particularly, for the three representative wavelengths analysed in this work ($\lambda_L = 2.2$ μm, $2.6$ μm, $3.2$ μm) the corresponding values of $N_e^{(m-Si)}$ are equal to $10.03\times10^{20}$ $cm^{-3}$, $8.91\times10^{20}$ $cm^{-3}$, $7.73\times10^{20}$ $cm^{-3}$, respectively. It is evident that metallization of Si with Mid-IR pulses occurs at significantly lower excitation conditions than at shorter wavelengths [7, 47, 53]. In particular, results shown in Fig.9a indicate that $N_e^{(m-Si)}$ decreases at increasing $\lambda_L$.

In regard to the possibility to metallise part of the Si substrate and turn it to a plasmonic material, without loss of generality, our analysis is centred on the investigation of the irradiation a $SiO_2$/Si stack at $F$=0.18 J/cm$^2$, $F$=0.20 J/cm$^2$ and $F$=0.24 J/cm$^2$ for $\lambda_L = 3.2$ μm, $\lambda_L = 2.6$ μm and $\lambda_L = 2.2$ μm, respectively. The selection of these fluence values was to ensure that a thickness of a maximum size equal to ~15-35 nm of metallic Si is produced for all $d$ (see Supplementary Material) and therefore evaluate the impact of the metallic layer on the features of the induced plasmons. A similar procedure can be followed to deduce the impact of thicker $d_{(m-Si)}$ produced in different conditions, however, the objective of the present investigation is to reveal a potential influence of the $SiO_2$ coating**.**

In contrast to standard multilayered structures comprising only unexcited dielectric materials and metals, the evaluation of the dielectric parameters that is required for the solution of the dispersion relation is performed considering that the dielectric constant of both the non-metallic and metallic Si are carrier density dependent (air and $SiO_2$ are considered unexcited and their refractive index is considered constant throughout the corresponding regions); the following assumptions are made for the sake of simplicity and without loss of generality:

(i) the carrier density values (i.e. inside the depth of the excited solids) resulting after the intensity reaches the maximum value will be used to calculate the dielectric parameters,

(ii) as there is a variation of the values across the depth of the excited solids, a mean value of the dielectric parameters will be used for both metallic and non-metallic Si. In particularly, for the non-metallic Si, the mean value of the dielectric parameter in a ~15-35 nm is also used.

A numerical solution of the dispersion relation Eq.9 yields the SP wavelength $\Lambda$ for the three laser wavelengths $\lambda_L$ which are depicted in Fig.9b,c,d, respectively. Results show that for a single pulse, the SP wavelength drops fast at increasing $d$ (for $d$<1 μm) reaching a value that is significantly smaller than $\lambda_L$ (i.e. 30% to 35% smaller than $\lambda_L$). The physical mechanism can be understood from the stack configuration for $d\rightarrow0$ (i.e. absence of coating) and large $d$. In the former case, the SP on the coating/metallic Si tends to 'see' only the air and the air/metal dispersion relation will determine the SP periodicity that is of the size of $\lambda_L$ (i.e. $\sim \frac{\lambda_L}{Re\sqrt{\frac{\varepsilon_{air}\varepsilon_{metallic\,Si}}{\varepsilon_{air}+\varepsilon_{metallic\,Si}}}}$). By contrast, at large $d$ the dielectric/metal governs primarily the dispersion yielding periodicities approximately equal to $\sim \frac{\lambda_L}{Re\sqrt{\frac{\varepsilon_{SiO_2}\varepsilon_{metallic\,Si}}{\varepsilon_{SiO_2}+\varepsilon_{metallic\,Si}}}}$. In that case, $\Lambda^{\lambda_L=2.2\,\mu m}$~1.56 μm, $\Lambda^{\lambda_L=2.6\,\mu m}$~1.86 μm, and $\Lambda^{\lambda_L=3.2\,\mu m}$~2.25 μm. Similar predictions could also result following a more rigorous approach via the use of the dispersion relations for a three-layered stack, 'air/(metallic Si)/Si' or 'dielectric coating/(metallic Si)/Si' [20, 50, 54, 55]

$$1 = \frac{k_3/\varepsilon_3 + k_1/\varepsilon_1}{k_3/\varepsilon_3 - k_1/\varepsilon_1} \times \frac{k_3/\varepsilon_3 + k_4/\varepsilon_4}{k_3/\varepsilon_3 - k_4/\varepsilon_4} \quad \text{(absence of SiO}_2\text{ coating)} \tag{10}$$

$$0 \cong \frac{k_3/\varepsilon_3 + k_2/\varepsilon_2}{k_3/\varepsilon_3 - k_2/\varepsilon_2} \times \frac{k_3/\varepsilon_3 + k_4/\varepsilon_4}{k_3/\varepsilon_3 - k_4/\varepsilon_4} \quad \text{(very thick SiO}_2\text{ coating)} \tag{11}$$

should be used for $d$=0 and $d\rightarrow\infty$, respectively. Simulation results show that the competition between the dielectric coating and air leads to a gradual reduction of the SP periodicity that is completed at $d_{crit}$~1 μm. Thus, $d_{crit}$ represents a critical value of the $SiO_2$ thickness at which the coating can be assumed to be semi-infinite.

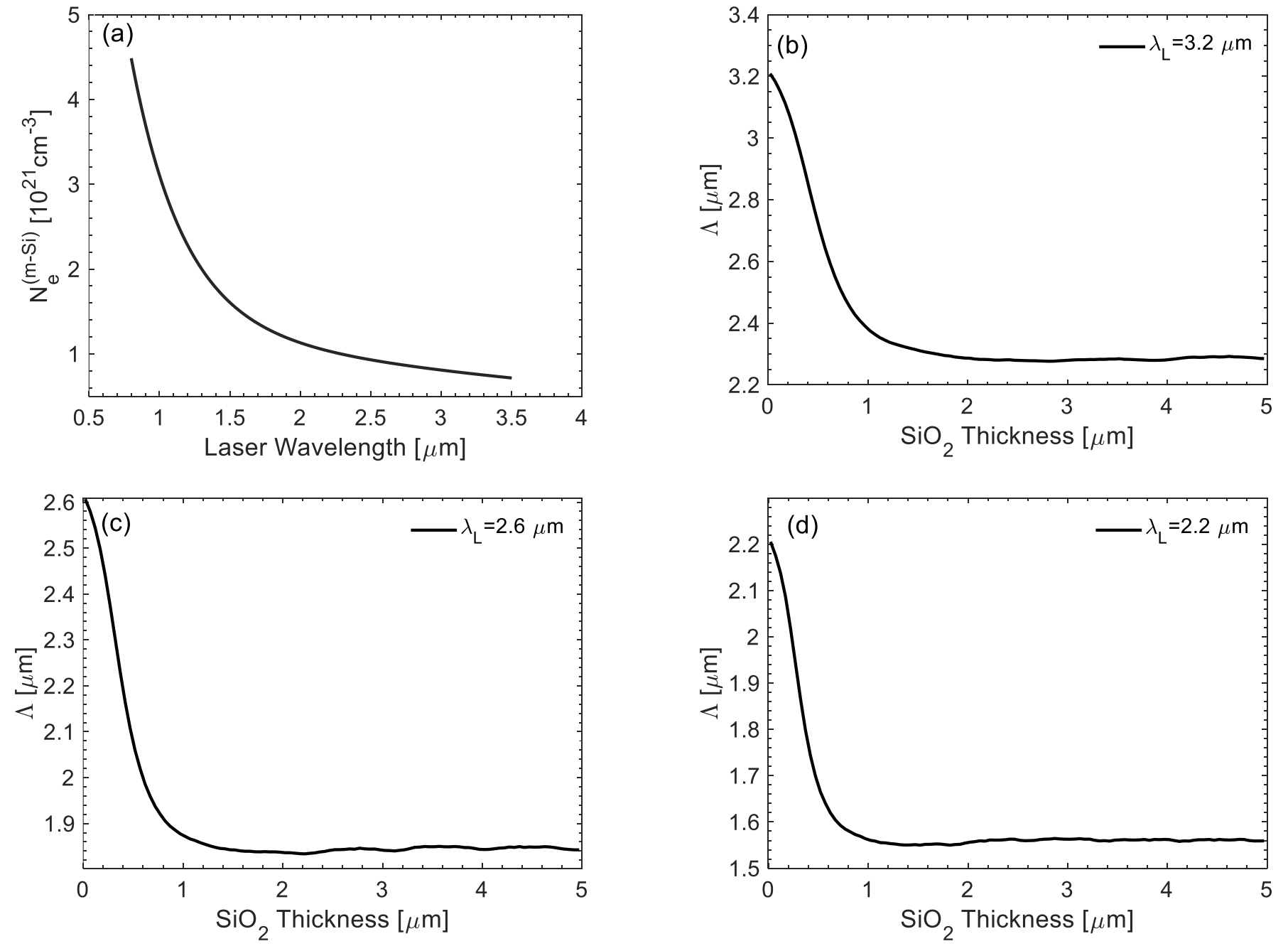

Figure 9: (a) Minimum Carrier density for metallisation of Si at various wavelengths. SP periodicity on the $SiO_2$/Si as a function of the $SiO_2$ thickness at (b) $\lambda_L$=3.2 μm, (c) 2.6 μm, and (d) 2.2 μm.

The energy of the SP is confined on the interface between the dielectric and the metallised Si and decays as an evanescent wave away from the interface. To evaluate, further to the behaviour of the SP, it is important to explore the spatial features of the electromagnetic modes that are excited on the dielectric/metal surface; these are (i) the damping length $L$ of the SP propagation along the surface and (ii) the decay length of the SP $L_D$ away from the interface [9, 56]. The damping (propagation) length represents the distance over which the intensity of the SP falls to *1/e* of its initial value and it is equal to $1/(2Im(\tilde{\beta}))$ [57]. As reported in a previous work [9], the capacity to modulate the SP propagation length is a crucial in laser-based processing of larger areas. Simulation results illustrated in Fig.10a indicate that both the laser wavelength and the dielectric coating thickness can be used to tailor the propagation length of the excited SP. A large drop for $d$<~1 μm at all wavelengths (~70% reduction) is predicted (Fig.10a) starting from values close to $L_{no\,coating}^{\lambda_L=2.2\,\mu m}$~17 μm, $L_{no\,coating}^{\lambda_L=2.6\,\mu m}$~25 μm, $L_{no\,coating}^{\lambda_L=3.2\,\mu m}$~38 μm that agree with the calculations from Eq.10. By contrast, as the thickness of the coating increases beyond $d$~1 μm, a small variation around a mean value is manifested while a maximum attained value is similar (Fig.10a) to the predictions derived from Eq.11 ($L_{max}^{\lambda_L=2.2\,\mu m}$~6 μm, $L_{max}^{\lambda_L=2.6\,\mu m}$~10.5 μm, and $L_{max}^{\lambda_L=3.2\,\mu m}$~17 μm) for large $SiO_2$ thicknesses. Thus, a significant modulation of the propagation length of the electromagnetic modes is possible via appropriate selection of the thickness of the dielectric coating.

In regard to the decay of the SP, $L_D^{(metallic\ Si)}$ represents the SP confinement *inside* the metallic material and it is associated with the (skin) depth at which the SP field amplitude falls to *1/e* (i.e. the field amplitude decreases exponentially as $e^{-k_3|z|}$), normal to the metallic surface surface [56, 58] and thus, $L_D^{(metallic\ Si)}$ $=1/k_3$. Similarly, $L_D^{(SiO_2)}$ $=1/k_2$ provides an estimate of the (skin) depth at which the SP field amplitude falls to *1/e inside* the dielectric material. Finally, $L_D^{(Si)}$ $=1/k_4$ provides an estimate of the (skin) depth at which the SP field amplitude falls to *1/e inside* the non-metallic volume of Si.

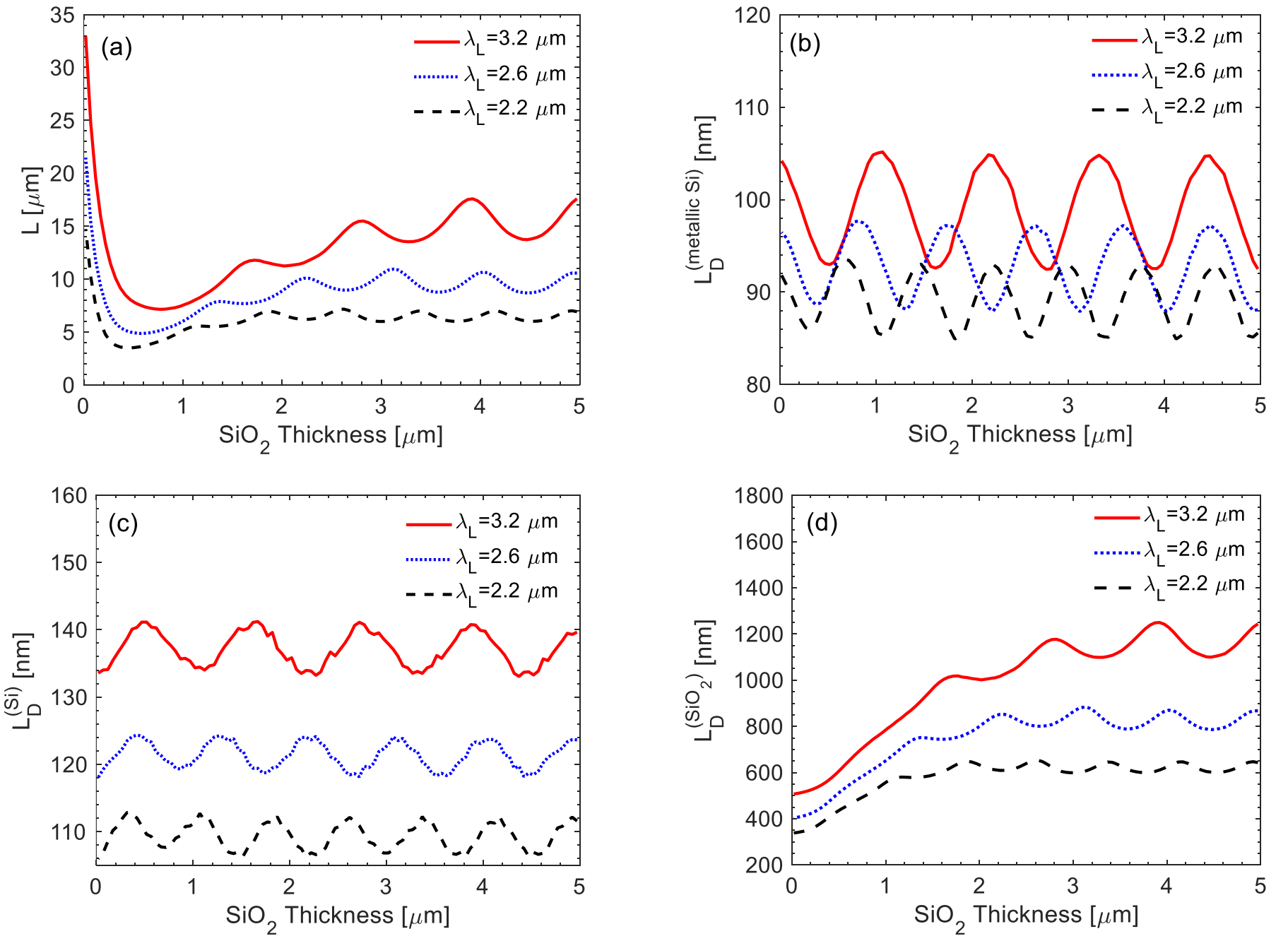

Figure 10: SP (a) propagation length and the decay length *inside* (b) metallic Si, (c) non-metallic Si and (d) $SiO_2$ as a function of the $SiO_2$ thickness at various laser wavelength ($\lambda_L$=2.2 μm, 2.6 μm, 3.2 μm). *Results are shown for d between 20 nm and 5 μm.*

Simulations illustrated in Fig.10b for $L_D^{(metallic\ Si)}$ show a small variation (~5%) around a mean value at all three wavelengths. In particularly, the mean value is equal to $L_D^{(metallic\ Si)\ \lambda_L=2.2\ \mu m}$ = 89 nm, $L_D^{(metallic\ Si)\ \lambda_L=2.6\ \mu m}$ = 93 nm and $L_D^{(metallic\ Si)\ \lambda_L=3.2\ \mu m}$ = 99 nm which are similar to the values in absence of the coating and large *d* produced from Eq.10 and Eq.11, respectively. A similar behaviour is exhibited for $L_D^{(Si)}$ (Fig.10c) for which a small variation (~10%) of the decay length of the SP inside the non-metallic Si occurs around $L_D^{(Si)\ \lambda_L=2.2\ \mu m}$ = 109 nm, $L_D^{(Si)\ \lambda_L=2.6\ \mu m}$ = 121 nm and $L_D^{(Si)\ \lambda_L=3.2\ \mu m}$ = 137 nm for the three wavelengths which is also confirmed by the use of Eqs.10, 11. Our results show that although the laser wavelength can be used to control the SP confinement inside the metallic/non-metallic Si (i.e. larger confinement is predicted at smaller wavelengths) the coating thickness does not appear to induce remarkable changes to the decay length in those regions. Similar insignificant variations are also calculated at different fluences (results not shown) where higher excitation levels are expected. By contrast, a remarkable increase of $L_D^{(SiO_2)}$ with the coating thickness is predicted. More specifically, results illustrated in Fig.10d show a large increase of the decay length inside the coating with the thickness before reaching a maximum value that is provided by Eq.11 for *d*>1.5 μm-2 μm. More specifically, simulations predict $L_D^{(SiO_2)\ \lambda_L=2.2\ \mu m}$ = 620 nm, $L_D^{(SiO_2)\ \lambda_L=2.6\ \mu m}$ = 820 nm, and $L_D^{(SiO_2)\ \lambda_L=3.2\ \mu m}$ = 1150 nm. Similarly, the variation predicted for large values does not appear to be influenced by *d*.

Interesting results also are revealed for the energy of the SP confined to the interface of the dielectric/metallic Si. The *absolute value* (magnitude) of the magnetic field distribution $|H_y|$ is shown in Fig.11 (analytic expressions for $H_y$ are provided in the Supplementary Material) offers a measure of the strength and confinement of the SP mode for $\lambda_L$=2.2 μm (Fig.11a,b,c), for $\lambda_L$=2.6 μm (Fig.11d,e,f), and $\lambda_L$=3.2 μm (Fig.11g,h,i) for three values of the coating thickness, (*d*=70 nm, *d*=470 nm and *d*=1.47 μm) for each laser wavelength. It is noted that $|H_y|$ values are normalized to 1. Similar conclusions can be deduced for different thicknesses and laser conditions. Results, not only, indicate the distinct SP features discussed above, but also, the strength of the confinement of the SP mode that depends on the coating thickness that is used. On the other hand, a different decay length inside $SiO_2$ is deduced. It is shown that at the same laser wavelength, the confinement of the SP mode inside $SiO_2$ can be controlled via appropriate modulation of the coating thickness at each wavelength (Fig.11). These simulations and the aforementioned discussion on $L_D^{(SiO_2)}$ show that for small coating thicknesses, at increasing $d$, $L_D^{(SiO_2)}$ increases while it exceeds the size of the dielectric (Fig.10d); this behaviour indicates that the evanescent wave decays inside the air (Fig.11) and this effect appears to be more pronounced at longer laser wavelengths. Thus, localization of the plasmonic modes in the $SiO_2$/metallic Si increases at smaller wavelengths. Therefore, for longer wavelengths (Fig.11), while most of the energy of the transversely confined guided modes is stored in the coating, there is an amount of the energy of the mode that is stored in air in contrast to the behaviour at shorter wavelengths where the mode 'sees' less the outer dielectric. On the other hand, the behaviour of the propagation length of the SP field along the *x*-axis discussed above (Fig.10a) is illustrated in Fig.11: the damping depth increases with the laser wavelength while at each wavelength a decrease of $L$ (for $d$=0.47 μm) is followed by an increase of the propagation length for thicker coatings ($d$=1.47 μm).

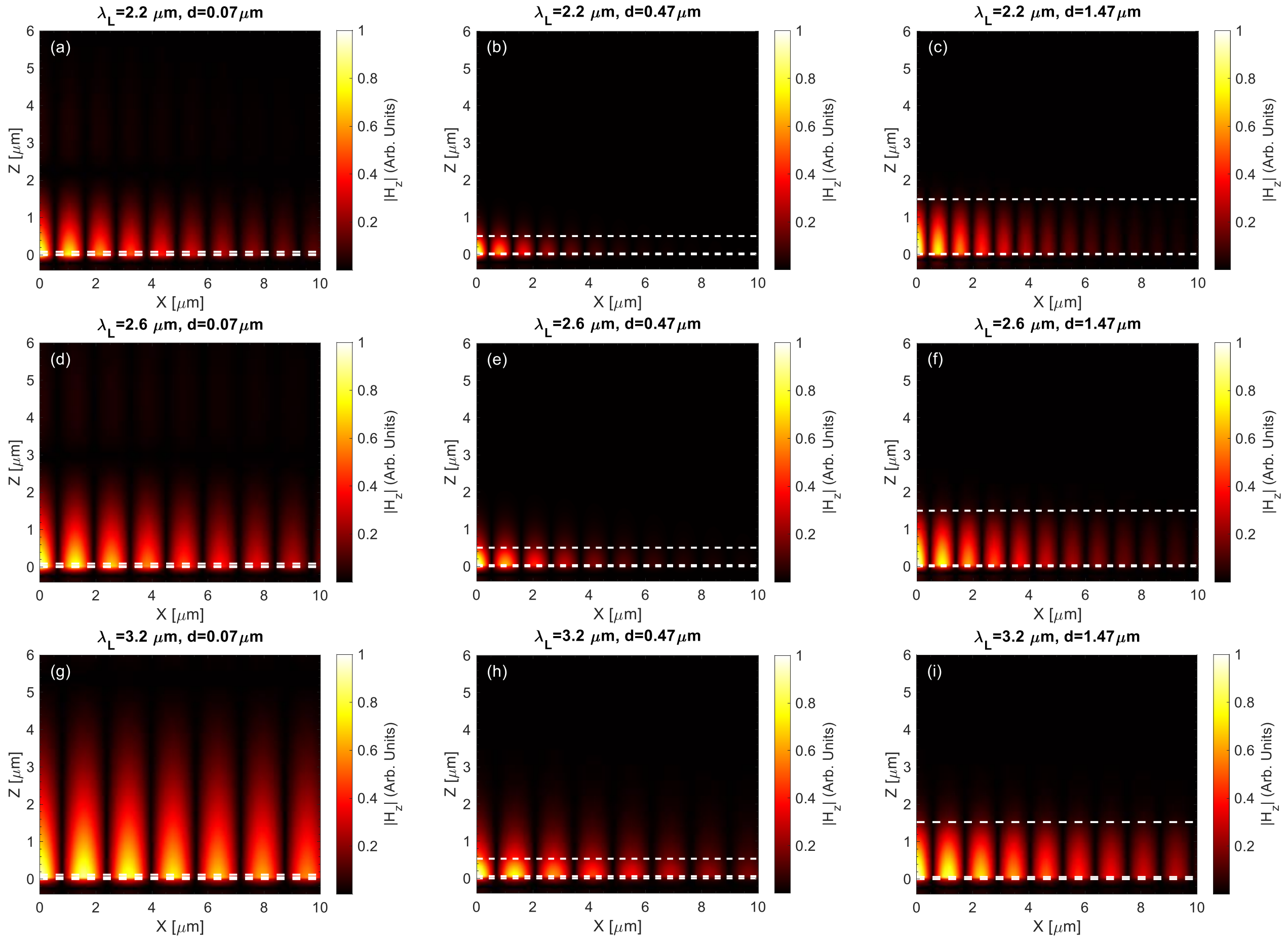

Figure 11: Absolute value of the magnetic field distribution ($|H_y|$) for various values of $d$ at $\lambda_L$=2.2 μm (a, b, c), $\lambda_L$=2.6 μm (d, e, f), $\lambda_L$=3.2 μm (g, h, i). Upper and lower *white* dashed lines indicate the position of the interface between air/$SiO_2$, $SiO_2$/metallic Si, respectively. Three coatings of $d$=70 nm, $d$=470 nm and $d$=1.47 μm were considered. Results are normalized to 1.

The above discussion of the dependence of the SP propagation and decay lengths on $d$ for thin and thick coatings indicates the significant role of the dielectric film in the behavior of the SP. The results demonstrate the interplay between air and the dielectric in the determination of the dispersion of an air/$SiO_2$/metallic Si stack. For thick materials, the SP mode does not 'feel' significantly the outside dielectric (air). By contrast, as the coating becomes thinner, the dispersion is determined by both the dispersion of air/metallic Si and $SiO_2$/metallic Si. Finally, in absence of the fused silica coating, the air/metal dispersion dictates the SP features. Thus, tailoring the dispersion relations of the SPs is feasible via the employment of an air/$SiO_2$/(metallic Si) stack. The above analysis demonstrates that the presence of two dielectric layers (air and $SiO_2$) offer a significantly higher flexibility to modulate the dispersion relation than with only one dielectric layer [9]. A similar analysis could be extended to a multilayered stack comprising of more than two dielectric materials on top of a metallic substrate, however, this is beyond the scope of the current work. Nevertheless, it has been shown that the dispersion curve can be systematically controlled by changing the dielectric material thicknesses and their refractive indices [59].

The above predictions as well as the SP features discussed above could be exploited for potential applications in photonics, plasmonics [2], waveguide fabrication [20] or nano/microfabrication of photonic components and optical signal processing devices at the sub-wavelength scale [9, 60].

## V. DISCUSSION

One of the main objectives in the development of high-power ultrafast laser systems and optical components is the increase of the DT of the irradiated material via the use of protective coatings [17, 61, 62] which can control the key factors that cause degradation in the performance of high-power lasers. Therefore, it is interesting to explore potential implications derived from the predicted opposite effect, a *decrease* of the fluence at which the onset of the damage of the substrate occurs via the presence of the coating of a lower refractive index than that of the substrate. To address this challenge, it is noted that Si modification with Mid-IR pulses via a nonlinear absorption is of particular significance for laser processing and laser-based applications [2, 4, 5]. In a previous report [9, 12], simulations showed that excitation of SP which account for the formation of periodic patterns on Si can be achieved at *lower* fluences at Mid-IR compared to shorter wavelengths (our results also showed that even a metallization of Si occurs at lower fluences). Thus, our predictions for the decrease of DT in the presence of the coating signifies that the presence of $SiO_2$ film can further facilitate (i) the onset of processing and (ii) excitation of SP electromagnetic modes on the $SiO_2$/Si interface at lower fluences and, therefore, allow surface modification of Si offering an innovative, greener route to laser-based manufacturing. Simulation results presented in the previous sections describe both the dependence of DT on the laser wavelength and how the $SiO_2$ thickness can empower the excitation of SP electromagnetic modes for a range of photon energies in the Mid-IR spectral region.

One significant aspect of the capacity to damage the substrate or allow fabrication of periodic patterns on Si via the employment of the coating is that it appears that the transparency of the upper film and ability to inscribe or (laser) write in the volume of the $SiO_2$/Si are capable to lead to some interesting potential applications (i.e. data storage [63]). In regard to the capacity to fabricate subwavelengths on the $SiO_2$/Si interface, not only the use of the dielectric material coating can allow the formation of subwavelength periodic structures on Si which are remarkably smaller than the laser wavelength but also, their periodicities can be modulated by ~30%-35% via appropriately selected $SiO_2$ thicknesses. Similarly, our results showed that other features of the induced SPs such as the SP decay inside the dielectric or the propagation length appear to be significantly dependent on the coating thickness and the laser wavelength. More specifically, our simulation results manifested that a remarkable variation in these parameters is expected. Thus, a potential influence of the $SiO_2$ thickness on the features of the electromagnetic modes such as the SP wavelengths, the decay length and confinement of the produced SP can be potentially exploited to present new capabilities in laser-based manufacturing, plasmonics and photonics. A more detailed analysis of the characteristics of the electromagnetic modes and the coupling of SP with the incident beam towards inducing morphological changes inside the $SiO_2$/Si could be the subject of a future investigation.

As the behaviour of the $SiO_2$ coating plays a crucial role in the opto/thermal response of the double-layered stack, it is important to ensure that the dielectric material possesses desirable properties. More specifically, in this work it has been assumed that the coating is transparent in the whole range of the irradiation wavelengths. It is known that despite the transparency of fused silica at lower wavelengths (for $\lambda_L < 1$ μm), the manufacturing process of the dielectric material influences its optical response at *longer* wavelengths. For example, the UV grade fused silica which is manufactured via the oxidation of high purity silicon by hydrolysis exhibits dips in the transmission spectrum centered at 1.4 μm, 2.2 μm, and 2.7 μm due to absorption from hydroxide (OH-) ion impurities [37, 38]. By contrast, IR grade fused silica (IR-FS) is characterized by a reduced amount of OH-ion impurities and transmission is high for up to 5 μm, however, again for $\lambda_L >$ 3.5 μm and depending on the thickness of IR-FS [64] absorption occurs. Nevertheless, it has been shown that the occurrence of the variability of the transmissivity is also significantly dependent on the fabrication process and on the thickness of the dielectric material [64]; therefore, a thorough investigation is required to evaluate the role of both the density of (OH-) ion impurities and $SiO_2$ thickness. Furthermore, an analysis of the impact of the $SiO_2$ thickness on the levels of excitation and the features of the produced SP modes are performed. For the sake of simplicity, IR-FS coatings are used in this study and it is assumed that the selected thickness of the films is appropriately small that maintains their high transmissivity (~1). To avoid potential discrepancies due to an enhanced absorptivity for $\lambda_L >$ 3.5 μm, the analysis is performed at wavelengths $\lambda_L <$ 3.5 μm. Nevertheless, the theoretical framework presented in this work can be extended to describe energy absorption from the coating.

The multireflection theory applied to calculate the optical parameters is assumed to provide an accurate methodology to determine the optical response of the irradiated stack. Although, still untested in a double-layered stack consisting of two transparent materials for Mid-IR pulses, this theory yields a satisfactory evaluation of the optical parameters for a variety of materials [65-67]. More complex approaches involving a detailed investigation of the electromagnetic wave propagation and the optical parameters evaluation for two-layered complex via the employment of a Maxwell solver coupled to electron dynamics could be the subject of a future study [68]. A future investigation can also involve the description of optical and electromagnetic phenomena and the ultrafast dynamics in a multilayered stack of transparent films of various thicknesses [18, 19].

To summarise, the theoretical predictions presented in this work manifest that a modulation of the optical parameters, ultrafast dynamics, damage threshold and features of excited SP are possible through an appropriate selection of the thickness of $SiO_2$. Interesting physical phenomena and dependencies appear to occur at various laser wavelengths and pulse durations assuming different thicknesses of the coating. The development of appropriate experimental protocols is certainly required to validate the theoretical model. Further investigation in different conditions such as shorter pulse durations or longer laser wavelengths will, also, allow to attain a more complete picture of the underlying physical mechanisms occurring following irradiation of multi-layered materials with Mid-IR fs pulses.

## VI. CONCLUSIONS

In this work, a detailed analysis of the ultrafast phenomena and thermal response of a $SiO_2$ coating on Si of various thicknesses of the coating following irradiation of the complex with ultrashort pulsed lasers in the Mid-IR range was presented. Results demonstrated a periodic variation of the optical parameters with the thickness of $SiO_2$ derived via the employment of multireflection theory. It was shown that the periodic behavior was projected into the ultrafast dynamics of the excited carriers and thermal response of the substrate while no part of laser energy is absorbed from the coating itself. A remarkable impact of the coating thickness was manifested and it can be used to modulate the damage threshold of Si while it was shown that this parameter can decrease the threshold by up to ~30% (depending on the pulse width) in comparison to the value in absence of the coating. Similar conclusions were deduced for various fluences, pulse durations and laser wavelengths while the pronounced role of the absorbed energy from the substrate on the damage threshold was emphasised. Finally, it was shown that the coating thickness can be exploited to control the features of the excited SP modes (leading to a ~30%-35% variation in the SP periodicity, propagation (~70%) and decay lengths compared to predictions in absence of the coating) with a potential benefit to applications in photonics, plasmonics, waveguide fabrication or nano/microfabrication. The elucidation of the laser driven physical phenomena that characterize the response of optical coatings on materials with Mid-IR fs pulses via this analysis and the remarkable predictions can be employed for the development of new tools for nonlinear optics and photonics for a large range of Mid-IR laser-based applications.

## ACKNOWLEDGEMENTS

The authors would like to acknowledge financial support from *HELLAS-CH* project (MIS 5002735), implemented under the "Action for Strengthening Research and Innovation Infrastructures" funded by the Operational Programme "Competitiveness, Entrepreneurship and Innovation" and co-financed by Greece and the EU (European Regional Development Fund).

Corresponding authors:
♣ George D.Tsibidis: tsibidis@iesl.forth.gr (Theory and Simulations)
* Emmanuel Stratakis: stratak@iesl.forth.gr

# Supplementary Material

## 1. Transmissivity and Impact Ionisation contribution

Simulations are illustrated in Fig.SM1 to describe the role of the laser wavelength on the transmissivity (Fig.SM1a,b,c), carrier density (Fig.SM1d,e,f) into the substrate as well as the influence of the impact ionisation (Fig.SM1d,e,f) following irradiation of $SiO_2$/Si for three values of the coating thickness ($d$=20 nm, 720 nm, 1220 nm). The '*C*' symbol in the legend in the second row in Fig. SM1 stands for results derived assuming the complete model presented in the main manuscript while the rest of the lines were produced by neglecting the impact ionisation (in principle, assuming a multiphoton ionisation process).

Results manifest that the inclusion of the impact ionisation processes simply increase the carrier density but it does not change the wavelength-dependent carrier density order at any coating thickness (Fig.SM1d,e,f).

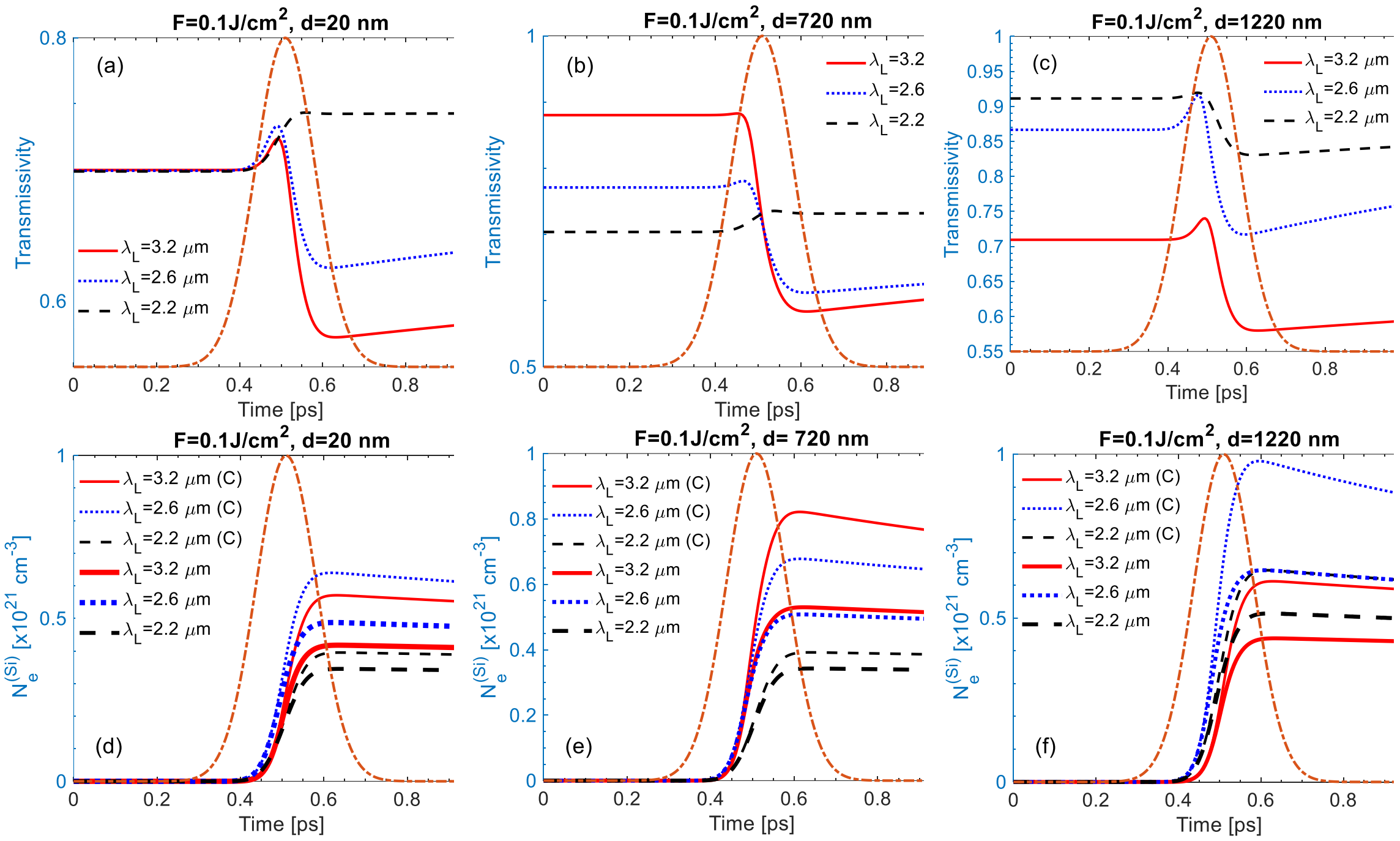

Figure SM1: Evolution of transmissivity (a,b,c) and Carrier lattice temperature (d,e,f) on the Si surface ($F$=0.1 J/cm$^2$, $\tau_p$=170 fs) for three $SiO_2$ thickness ($d$=20 nm,720 nm, 1220 nm) at various laser wavelength ($\lambda_L$=2.2 μm, 2.6 μm, 3.2 μm). The letter '*C*' in parentheses corresponds to the use of the complete model for describing the ultrafast dynamics.

## 2. Reflectivity and Transmissivity

Simulations are shown in Fig.S2 to illustrate the evolution of the reflectivity and the transmissivity. Results are depicted for $d$=520 nm at $F$=0.1 J/cm$^2$, $\tau_p$=170 fs, $\lambda_L$=3.2 µm. Similar results (not shown) have been derived in other conditions used in this work.

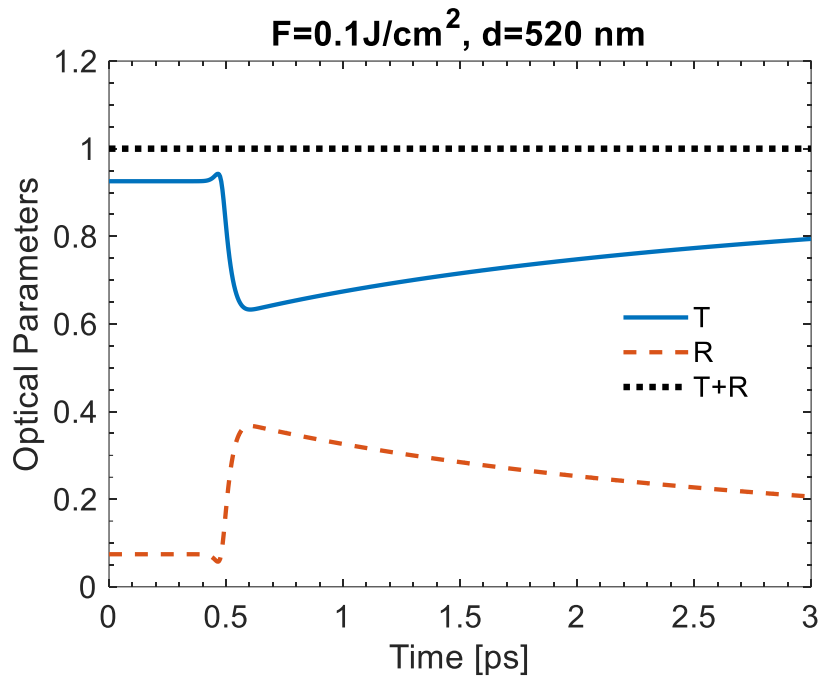

Figure SM2: Evolution of reflectivity and transmissivity at $F$=0.1 J/cm$^2$, $\tau_p$=170 fs for $d$=520 nm at $\lambda_L$=3.2 µm.

## 3. SP Electromagnetic modes

In a four-layered stack (Fig.SM3a), the dispersion relation is provided by the following expression

$$e^{-2k_2 d} = \frac{\left(1+\frac{k_3/\varepsilon_3}{k_2/\varepsilon_2}\right)\left(1+\frac{k_4/\varepsilon_4}{k_3/\varepsilon_3}\right)+\left(1-\frac{k_3/\varepsilon_3}{k_2/\varepsilon_2}\right)\left(1-\frac{k_4/_4}{k_3/\varepsilon_3}\right)e^{-2k_3 d_3}}{\left(1+\frac{k_4/\varepsilon_4}{k_3/\varepsilon_3}\right)\left(1-\frac{k_3/\varepsilon_3}{k_2/\varepsilon_2}\right)+\left(1-\frac{k_4/\varepsilon_4}{k_3/\varepsilon_3}\right)\left(1+\frac{k_3/\varepsilon_3}{k_2/\varepsilon_2}\right)e^{-2k_3 d_3}} \times \frac{\left(1+\frac{k_2/\varepsilon_2}{k_1/\varepsilon_1}\right)}{\left(-1+\frac{k_2/\varepsilon_2}{k_1/\varepsilon_1}\right)} \quad \text{(SM.1)}$$

where $k_j = \sqrt{\tilde{\beta}^2 - \varepsilon_j k_0^2}$ ($j$=1, 2, 3, 4 for 'air', '$SiO_2$', 'metallic Si', and 'Si'), $\varepsilon_j$ stands for the dielectric permittivity of the $j$ material, $d_3$ is the thickness of the layer of the 'metallic Si', $k_0$ (=2π/$\lambda_L$) is the free-space wavenumber at the laser wavelength and $\tilde{\beta}$ is the propagation constant of the SP. On the other hand, $d$ stands for the coating thickness.

In Fig.SM3b, the thickness of the metallic layer of Si is shown as a function of the $SiO_2$ thickness following excitation of the $SiO_2$/Si structure with $F$=0.18 J/cm$^2$, $F$=0.20 J/cm$^2$ and $F$=0.24 J/cm$^2$ for $\lambda_L$ = 3.2 µm, $\lambda_L$ = 2.6 µm and $\lambda_L$ = 2.2 µm, respectively.

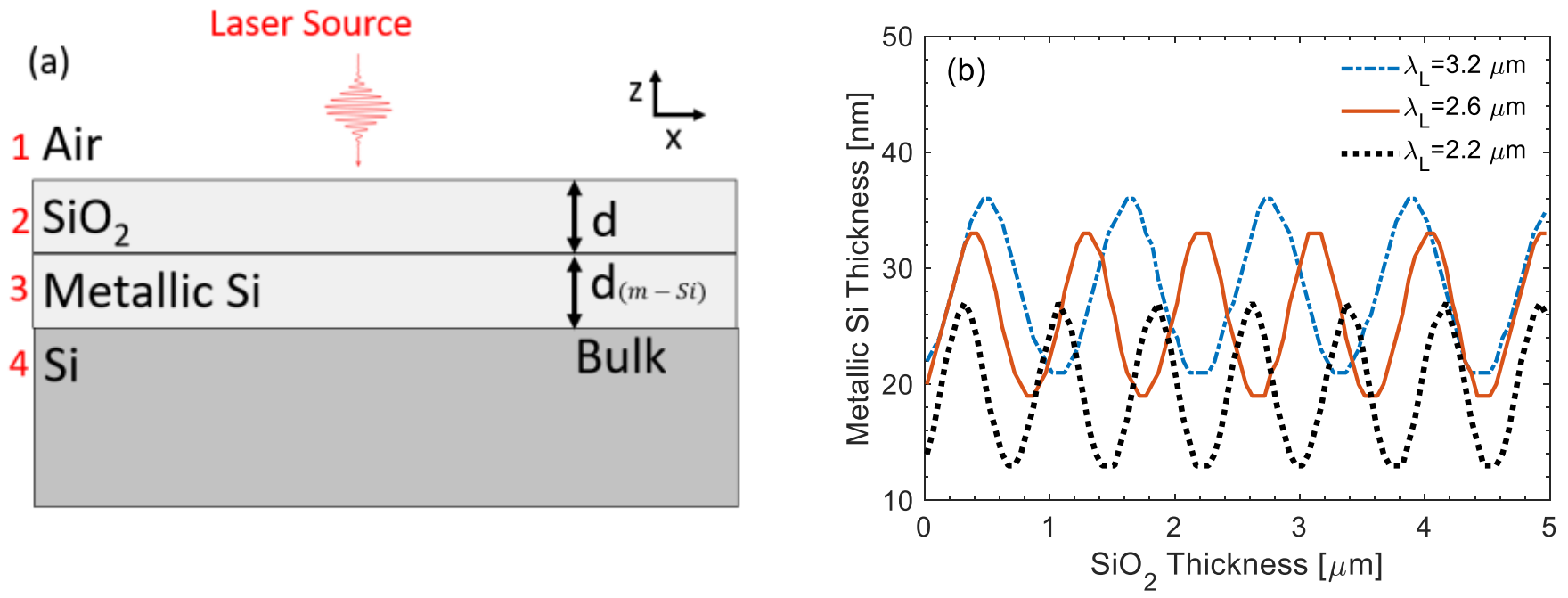

Figure SM3: (a) Four-layered stack, (b) Metallic Silicon thickness.

The magnetic field $H_y$ in 'air', '$SiO_2$', 'metallic Si', 'Si' is provided by the following expressions, respectively ($z$=0 is taken on the interface between metallic Si and Si)

$$
\begin{aligned}
H_y^{air}(x,z) &= Ae^{-k_1 z}e^{i\tilde{\beta}x} \\
H_y^{SiO_2}(x,z) &= \left[Be^{-k_2 z} + Ce^{k_2 z}\right]e^{i\tilde{\beta}x} \\
H_y^{metallic\ Si}(x,z) &= \left[De^{-k_3 z} + Ee^{k_3 z}\right]e^{i\tilde{\beta}x} \\
H_y^{Si}(x,z) &= e^{k_4 z}e^{i\tilde{\beta}x}
\end{aligned}
\qquad \text{(SM.2)}
$$

where

$$
\begin{aligned}
A &= \left(Be^{(k_2+k_1)(d+d_3)} + Ce^{(-k_2+k_1)(d+d_3)}\right)/4 \\
B &= \left[(1+K_{43})(1+K_{32})e^{(-k_2+k_3)d_3} + (1-K_{43})(1-K_{32})e^{-(k_2+k_3)d_3}\right]/4 \\
C &= \left[(1+K_{43})(1-K_{32})e^{(k_2+k_3)d_3} + (1-K_{43})(1+K_{32})e^{(k_2-k_3)d_3}\right]/4 \\
D &= (1+K_{43})/2 \\
E &= (1-K_{43})/2 \\
K_{32} &= 1+\frac{k_3\varepsilon_2}{k_2\varepsilon_3} \\
K_{43} &= 1+\frac{k_4\varepsilon_3}{k_3\varepsilon_4}
\end{aligned}
\qquad \text{(SM.3)}
$$